\documentclass[aps,twocolumn,showpacs,amsmath,superscriptaddress,amssymb,prc]{revtex4-2}
\usepackage{graphicx,here}
\usepackage{multirow}
\usepackage{tikz}
\usepackage{epstopdf}
\usepackage[percent]{overpic}
\usepackage{scrextend}
\usepackage{xcolor,xspace}
\usepackage[ulem=normalem]{changes}
\usepackage{hyperref}
\hypersetup{colorlinks=True,urlcolor=blue,linkcolor=blue,citecolor=blue,filecolor=black}

\colorlet{Changes@Color}{red}
\usepackage{dcolumn,color,footnote,bm,braket}
\usepackage{url,longtable,tabularx}
\usepackage{threeparttable}

\usepackage{fancybox}
\usetikzlibrary{matrix}

\begin{document}

\title{Signatures of shape phase transitions in krypton isotopes based on relativistic energy density functionals}

\author{K.~E.~Karakatsanis}
\affiliation{Department of Physics, Faculty of Science, 
University of Zagreb, HR-10000 Zagreb, Croatia}
\affiliation{Physics Department, Aristotle University of Thessaloniki, Thessaloniki GR-54124, Greece}

\author{K.~Nomura}
\email{knomura@phy.hr}
\affiliation{Department of Physics, Faculty of Science, 
University of Zagreb, HR-10000 Zagreb, Croatia}

\date{\today}

\begin{abstract}
Spectroscopic properties that characterize the shape phase transitions 
in krypton isotopes with the mass $A\approx80$ region are 
investigated 
within the framework of the nuclear density functional theory. 
Triaxial quadrupole constrained self-consistent mean-field 
calculations that employ relativistic energy density functionals 
and a pairing interaction are carried out for the even-even 
nuclei $^{76-86}$Kr. 
The spectroscopic properties are computed by 
solving the triaxial quadrupole collective Hamiltonian, 
with the ingredients, i.e., 
the deformation-dependent moments of inertia and 
mass parameters, and the collective potential, 
determined by using the SCMF solutions as microscopic inputs. 
Systematic behaviors of 
the SCMF potential energy surfaces, 
the corresponding low-energy spectra, electric quadrupole and monopole 
transition probabilities, and the fluctuations in the triaxial 
quadrupole deformations  
indicate evolution of the underlying nuclear structure 
as functions of the neutron number,  
that is characterized by a considerable degree of shape mixing. 
A special attention is paid to the transitional nucleus $^{82}$Kr, 
which has been recently identified experimentally as an 
empirical realization of the E(5) critical-point symmetry. 
\end{abstract}

\maketitle

\section{Introduction}

Quantum phase transitions (QPTs) are prominent phenomena 
in many areas of physics and chemistry. In the atomic nucleus,  
a class of QPT is suggested to occur between different intrinsic 
shapes in the ground state 
\cite{cejnar2009,cejnar2010,carr-book,iachello2011a,cejnar2016,fortunato2021}, 
which has been empirically identified as an abrupt change of 
observables along a given isotopic or isotonic 
chain with the addition or subtraction of only a few nucleons. 
A typical example is the phase transition from vibrational 
to rotational energy spectra around the neutron number $N=90$ 
in the rare-earth region. 
The nuclear shape QPTs take place 
as functions of a discrete control parameter, i.e., nucleon number, 
hence an important question arises, as to how one can identify 
a particular nucleus 
as the critical point of the phase transitions. 
In addition, since the nuclei are finite quantum systems, 
the shape phase transitions are more or less smeared out 
in most of the realistic cases, which 
in turn points to another important question as to which physical 
observables can be regarded as the quantum order parameters 
of the phase transitions. 

In the language of the interacting boson model (IBM) \cite{IBM}, 
the nuclear shape QPTs can be interpreted in terms of the 
transitions between different dynamical symmetries that 
emerge from the bosonic algebras, 
i.e., U(5), SU(3), and O(6) limits, which are 
associated with the spherical vibrational, 
deformed rotational, and $\gamma$-unstable rotational states, 
respectively. 
A different class of symmetry, referred to as 
the critical-point symmetry (CPS) 
was introduced in Refs.~\cite{iachello2000,iachello2001}, 
which provides a criterion to classify and interpret the nature of 
the collective states in transitional regions. 
This symmetry consists in exact solutions of the 
geometric collective model that employs a potential 
appropriate for modeling the phase transition of interest. 
In particular, the E(5) CPS \cite{iachello2000}  
corresponds to the transitional nuclei between 
the U(5) and O(6) limits, 
and is obtained analytically 
by solving the collective Schr\"odinger equation with 
a flat-bottomed potential characteristic of the 
U(5)-O(6) phase transition. 
The first empirical evidence for the CPS was suggested in the 
nucleus $^{134}$Ba \cite{casten2000}, where the low-energy spectrum 
and selection rules of the electric quadrupole transitions 
exhibit patterns predicted by the exact E(5) CPS. 
Numerous experimental investigations have been made 
to identify further evidence for the shape QPTs 
and the corresponding CPSs in wider ranges of the nuclear 
mass table, thus aiming to clarify whether the shape QPTs 
are ubiquitous phenomena in nuclear many-body systems. 
(see, e.g., Refs.~\cite{casten2009,cejnar2010}, 
and references are therein).

More recently, experimental evidence for the E(5) CPS has been suggested 
for the nucleus $^{82}$Kr \cite{rajbanshi2021}. 
This would present a first empirical realization 
of the E(5) symmetry in the mass region $A\approx80$, 
and extend the region of the critical-point phenomena 
to lighter mass regions that were hitherto not 
as extensively pursued as in the case of heavier nuclei. 
Empirically, the mass $A\approx80$ nuclei around 
the Kr chain have also been suggested  
to demonstrate a rich variety of the nuclear structure 
phenomena, including the emergence of shape 
coexistence around the neutron sub-shell gap $N=40$ 
\cite{wood1992,clement2007,ljungvall2008,heyde2011}. 
Apart from the nuclear-structure point of view, 
the nucleus $^{82}$Kr is also of particular interest, 
since it corresponds to the final-state nucleus of the 
possible neutrinoless double-$\beta$ decay of $^{82}$Se, and 
an accurate theoretical calculation for its low-lying states 
is useful for a reliable prediction of the 
nuclear matrix element of this decay process.

On the theoretical side, the nuclear shape QPTs and the related 
spectroscopic properties have been extensively studied from 
various approaches, including the algebraic models 
\cite{cejnar2010,iachello2011a,fortunato2021,bonatsos2017,martinou2021}, 
the geometrical collective models \cite{cejnar2010,fortunato2021}, 
the large-scale shell model 
\cite{shimizu2001,caurier2005,togashi2016,shimizu2017}, 
and the methods based on the nuclear density functional theory (DFT) 
\cite{niksic2007,li2010,nomura2014,nomura2017odd-3,robledo2019,ebran2020}. 
In particular, the nuclear DFT framework has demonstrated an 
ability to provide an accurate, global, and computationally 
economical description of nuclear structure and dynamics \cite{schunck2019}. 
Both relativistic \cite{vretenar2005,niksic2011} 
and nonrelativistic \cite{bender2003,robledo2019} 
energy density functionals 
(EDFs) have been successfully applied in the 
global description of 
the bulk nuclear matter and ground-state properties, as well as 
collective excitations, over the entire region of the nuclear mass chart. 
The EDF framework is basically implemented in the  
self-consistent mean-field (SCMF) calculations \cite{RS}, 
in which an EDF is constructed as a functional of one-body 
nucleon density matrices that correspond to a single product state. 
To access spectroscopic properties, the EDF framework should 
be extended to take into account the dynamical correlations 
arising from the restoration of broken symmetries and fluctuations 
around the mean-field minima. 
A straightforward approach is the 
generator coordinate method (GCM) with symmetry projections 
and configuration mixing \cite{RS,bender2003,niksic2011,robledo2019}. 
In practical applications, the full GCM calculation in general 
becomes computationally demanding, especially in the case of 
heavy nuclei or when the large number of collective coordinates 
need to be taken into account. 
Alternative approaches to GCM have been provided, e.g., by the 
collective Hamiltonian \cite{niksic2009,prochniak2009,delaroche2010,niksic2011} 
and the mapped IBM \cite{nomura2008,nomura2010}. 
The EDF-based calculations both within the static and beyond 
SCMF approximations have also been extensively 
carried out to study the neutron-deficient Kr isotopes around 
the neutron sub-shell gap $N=40$ 
(see, e.g., Refs.~\cite{bender2006,girod2009,fu2013,trodriguez2014,yao2014,nomura2017kr,abusara2017,nomura2021qoch}).

Based on the relativistic EDF framework,  
here we investigate the evolution of the shape and 
low-lying states in the even-even Kr isotopes in the mass 
range $76\leq A \leq 86$, particularly focusing 
on the proposed E(5) CPS around the transitional nucleus $^{82}$Kr. 
The starting point is the triaxial quadrupole constrained 
SCMF calculations for the above Kr nuclei within the 
relativistic Hartree-Bogoliubov (RHB) framework 
\cite{vretenar2005,niksic2011}
using two representative classes of the relativistic EDF, 
i.e., the density-dependent meson-exchange (DD-ME2) \cite{DDME2}
and point-coupling (DD-PC1) \cite{DDPC1} interactions, 
and a separable pairing force of finite range \cite{tian2009}. 
Spectroscopic properties that can be considered signatures of 
the QPTs are computed by solving the 
collective Schr\"odinger equation with triaxial quadrupole 
shape degrees of freedom. The ingredients of the 
quadrupole collective Hamiltonian (QCH), that is, 
the deformation-dependent moments of inertia and mass parameters, 
and the collective potential, are determined by using the 
SCMF solutions as the microscopic inputs. 
Diagonalization of the QCH yields excitation spectra of low-energy 
positive-parity states and electric quadrupole and monopole 
transition rates. 
The RHB method that is combined with the QCH 
(denoted hereafter as RHB+QCH) has been employed in a 
number of previous theoretical investigations 
to predict and describe a variety of nuclear structure phenomena 
\cite{niksic2009,li2010,niksic2011,DDPC1-SHE,fu2013,li2016,xiang2018,prassa2021}. 

The paper is organized as follows. 
In Sec.~\ref{sec:theory}, we outline the RHB+QCH approach. 
In Sec.~\ref{sec:results} we present the SCMF results on the 
triaxial quadrupole potential energy surfaces, and the 
spectroscopic results on the low-energy excitation 
spectra, and $E2$ and $E0$ transition strengths. 
In the same section, we also show fluctuations of the $\beta$ 
and $\gamma$ deformations as another indicator of the 
phase transition. 
A special attention is given to the nucleus $^{82}$Kr, recently 
suggested to be an empirical realization of the E(5) symmetry, 
and a detailed comparison with the experimental and E(5) spectra 
is made. 
Finally, Sec.~\ref{sec:summary} gives a summary of the main results 
and conclusions.

\begin{figure*}
\begin{center}
\begin{tabular}{ccc}
\includegraphics[width=.32\linewidth]{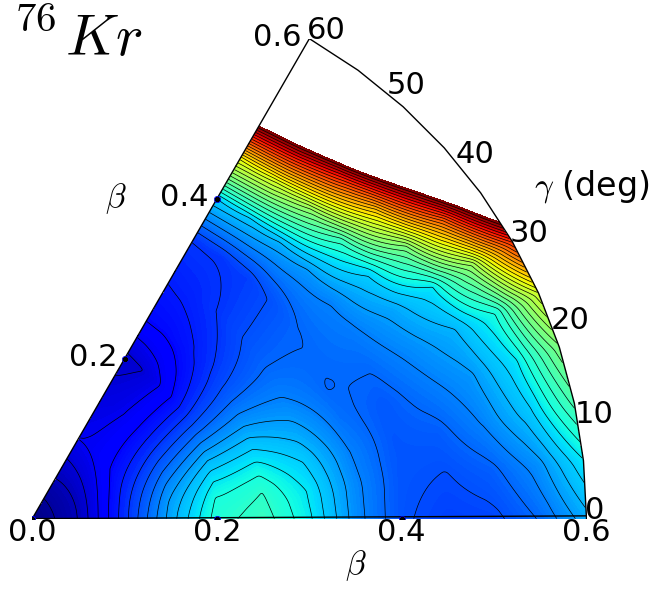} & 
\includegraphics[width=.32\linewidth]{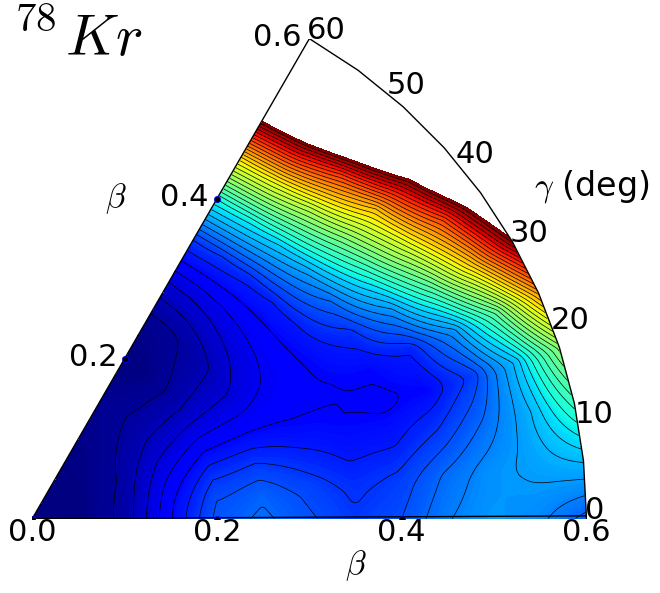} &
\includegraphics[width=.32\linewidth]{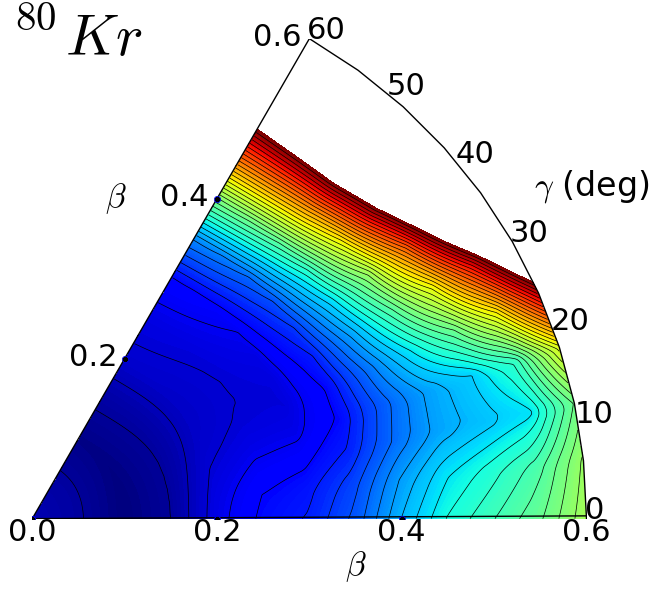} \\ 
\includegraphics[width=.32\linewidth]{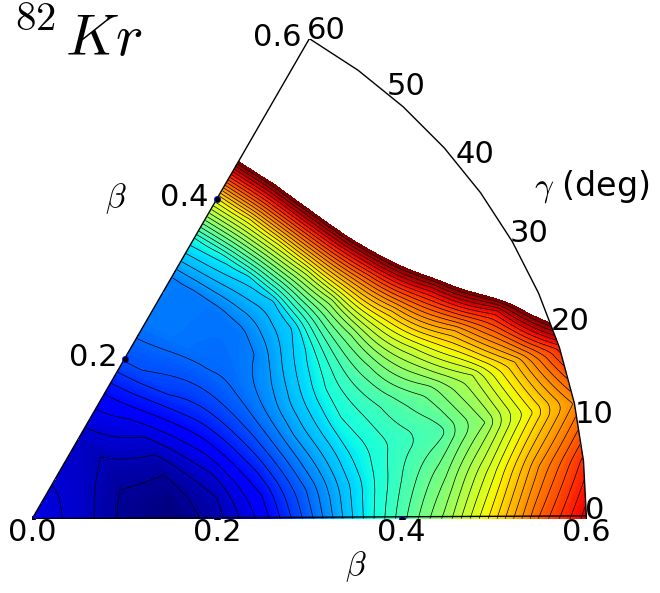} &
\includegraphics[width=.32\linewidth]{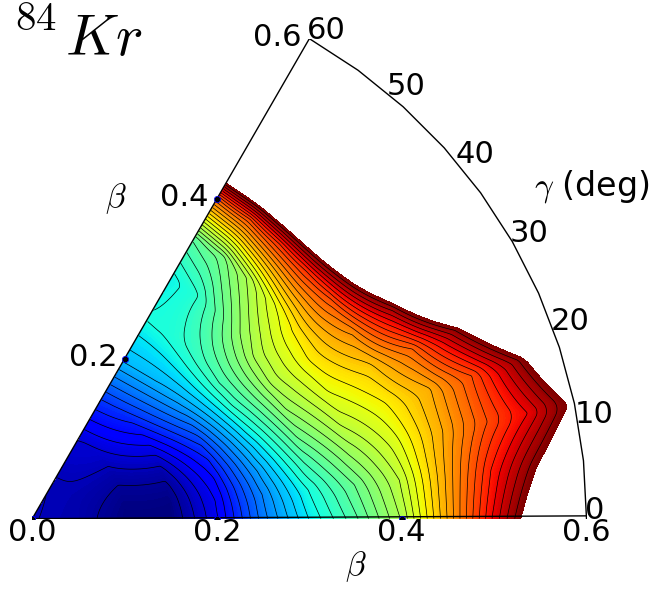} & 
\includegraphics[width=.32\linewidth]{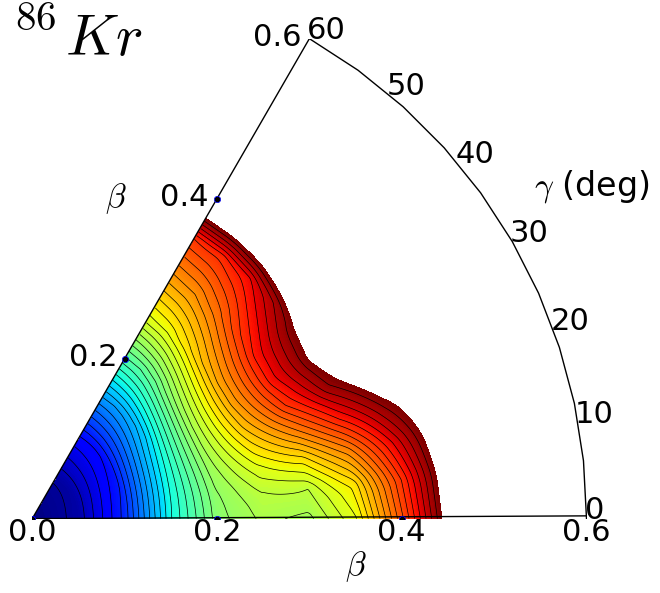} \\
\end{tabular}
\caption{Triaxial quadrupole potential energy surfaces 
for the even-even nuclei $^{76-86}$Kr in terms of the $\beta$ and 
$\gamma$ deformation variables, computed by the constrained 
SCMF calculations within the RHB framework based on the 
relativistic functional DD-ME2 and the separable pairing force 
of finite range. The total SCMF energies are plotted up to 10 MeV from the 
global minimum, and the energy difference between 
neighboring contours is 250 keV. 
}
\label{fig:pes}
\end{center}
\end{figure*}

\section{Theoretical framework\label{sec:theory}}

In this section, we give a brief description of the RHB+QCH 
approach adopted for the present theoretical analysis. 
For the detailed account of the formalism and numerical 
machinery in the constrained RHB framework, 
the reader is referred to 
Refs.~\cite{vretenar2005,niksic2011,DIRHB,DIRHBspeedup}, 
while the procedure to build the QCH from the SCMF solutions 
within the RHB is well documented, e.g., in 
Refs.~\cite{niksic2009,niksic2011}. 

The constraints imposed in the present SCMF calculations are 
on the expectation values of the mass quadrupole operators
\begin{align}
 \hat Q_{20}=2z^2-x^2-y^2 \; 
\quad\text{and}\quad
 \hat Q_{22}=x^2-y^2 \; ,
\end{align}
which are related to the 
axially-symmetric deformation $\beta$ 
and triaxiality $\gamma$ \cite{BM}, i.e., 
\begin{align}
\label{eq:bg}
& \beta=\sqrt{\frac{5}{16\pi}}\frac{4\pi}{3}\frac{1}{A(r_{0}A^{1/3})^{2}}
\sqrt{\braket{\hat{Q}_{20}}^{2}+2\braket{\hat{Q}_{22}}^{2}} \; , \\
& \gamma=\arctan{\sqrt{2}\frac{\braket{\hat{Q}_{22}}}{\braket{\hat{Q}_{20}}}} \; ,
\end{align}
with $r_0=1.2$ fm. The calculations are carried out 
in a harmonic oscillator basis, with the number of 
oscillator shells equal to 20. 
The separable pairing force of finite range, originally 
developed in Ref.~\cite{tian2009}, includes two sets of 
parameters that are determined so as to reproduce 
the pairing gaps resulting from the 
Gogny D1 and D1S effective interactions. 
Throughout this paper, the parametrization associated with 
the D1S force is employed, that is, the pairing strength 
$V_0=728$ MeV fm$^3$ and the parameter $a=0.644$ fm of the 
Gaussian function entering the separable interaction, 
for both the proton and neutron pairings. 
The constrained RHB calculations provide  
the potential energy surfaces (see Fig.~\ref{fig:pes}) 
and the SCMF single-particle solutions, 
which are subsequently used as the microscopic inputs to specify the 
collective Hamiltonian by the procedure described below. 

Quadrupole collective states are provided as the solutions of 
the triaxial QCH, with the deformation-dependent 
parameters determined by the constrained SCMF calculation  
within the RHB framework. The QCH is given by
\begin{align}
\label{eq:hamiltonian-quant}
\hat{H}_{\textnormal{coll}} 
= \hat{T}_{\textnormal{vib}}+\hat{T}_{\textnormal{rot}}
+V_{\textnormal{coll}} \; ,
\end{align}
with the vibrational kinetic energy:
\begin{align}
\hat{T}_{\textnormal{vib}} =
&-\frac{\hbar^2}{2\sqrt{wr}}
   \Biggl[\frac{1}{\beta^4}
   \Biggl(\frac{\partial}{\partial\beta}\sqrt{\frac{r}{w}}\beta^4
   B_{\gamma\gamma} \frac{\partial}{\partial\beta}
   \nonumber \\
&-\frac{\partial}{\partial\beta}\sqrt{\frac{r}{w}}\beta^3
   B_{\beta\gamma}\frac{\partial}{\partial\gamma}
   \Biggr)
   +\frac{1}{\beta\sin{3\gamma}}\Biggl(
   -\frac{\partial}{\partial\gamma} \sqrt{\frac{r}{w}}\sin{3\gamma}
\nonumber \\
&\times
B_{\beta \gamma}\frac{\partial}{\partial\beta}
    +\frac{1}{\beta}\frac{\partial}{\partial\gamma} \sqrt{\frac{r}{w}}\sin{3\gamma}
      B_{\beta \beta}\frac{\partial}{\partial\gamma}
   \Biggr)\Biggr] \; ,
\end{align}
and rotational kinetic energy:
\begin{align}
\label{eq:rot}
\hat{T}_{\textnormal{rot}} =
\frac{1}{2}\sum_{k=1}^3{\frac{\hat{J}^2_k}{\mathcal{I}_k}} \; ,
\end{align}
where $\hat{J}_k$ denotes the components of the angular momentum in
the body-fixed frame of a nucleus. 
The moments of inertia $\mathcal{I}_k$, 
as well as the mass parameters 
$B_{\beta\beta}$, $B_{\beta\gamma}$, and $B_{\gamma\gamma}$, 
depend on the quadrupole
deformation variables $\beta$ and $\gamma$ in such a way that 
$\mathcal{I}_k = 4B_k\beta^2\sin^2(\gamma-2k\pi/3)$. 
Two additional quantities that appear 
in the expression for the vibrational energy, i.e., 
$r=B_1B_2B_3$, and $w=B_{\beta\beta}B_{\gamma\gamma}-B_{\beta\gamma}^2 $,
determine the volume element in the collective space. 
The moments of inertia are
computed using the Inglis-Belyaev (IB) formula 
\cite{inglis1956,belyaev1961}, and the mass
parameters associated with the two quadrupole collective coordinates
$q_0=\langle\hat{Q}_{20}\rangle$ and $q_2=\langle\hat{Q}_{22}\rangle$
are calculated in the cranking approximation. 
The collective potential $V_{\textnormal{coll}}$ 
in Eq.~(\ref{eq:hamiltonian-quant}) is obtained by subtracting 
the zero-point energy corrections from the total RHB deformation energy.

The corresponding eigenvalue problem is solved using 
an expansion of eigenfunctions in terms
of a complete set of basis functions that depend on the 
deformation variables $\beta$ and
$\gamma$, and the Euler angles $\Omega=(\phi,\theta,\psi)$. 
The diagonalization of the Hamiltonian yields the excitation 
energies and collective
wave functions for each value of the total angular 
momentum and parity, that are used to calculate observables. 
Note that the present QCH approach is limited up to those 
spins at which the first band crossing takes place. 
The higher-spin states can be described by alternative 
approaches, e.g., by cranking models. Illustrative examples 
are found in Ref.~\cite{afanasjev2005}, dealing with 
the $^{72,74,76}$Kr nuclei. 

An important advantage of using the collective model 
based on SCMF single-(quasi)particle 
solutions is the fact that physical observables, 
such as transition probabilities and spectroscopic quadrupole 
moments, are calculated in the full configuration space and
there is no need for effective charges. 
Using the bare value of the proton charge in the
electric quadrupole operator, the transition probabilities 
between eigenvectors of the collective Hamiltonian 
can be directly compared with data.

\begin{figure*}[ht]
\begin{center}
\includegraphics[width=\linewidth]{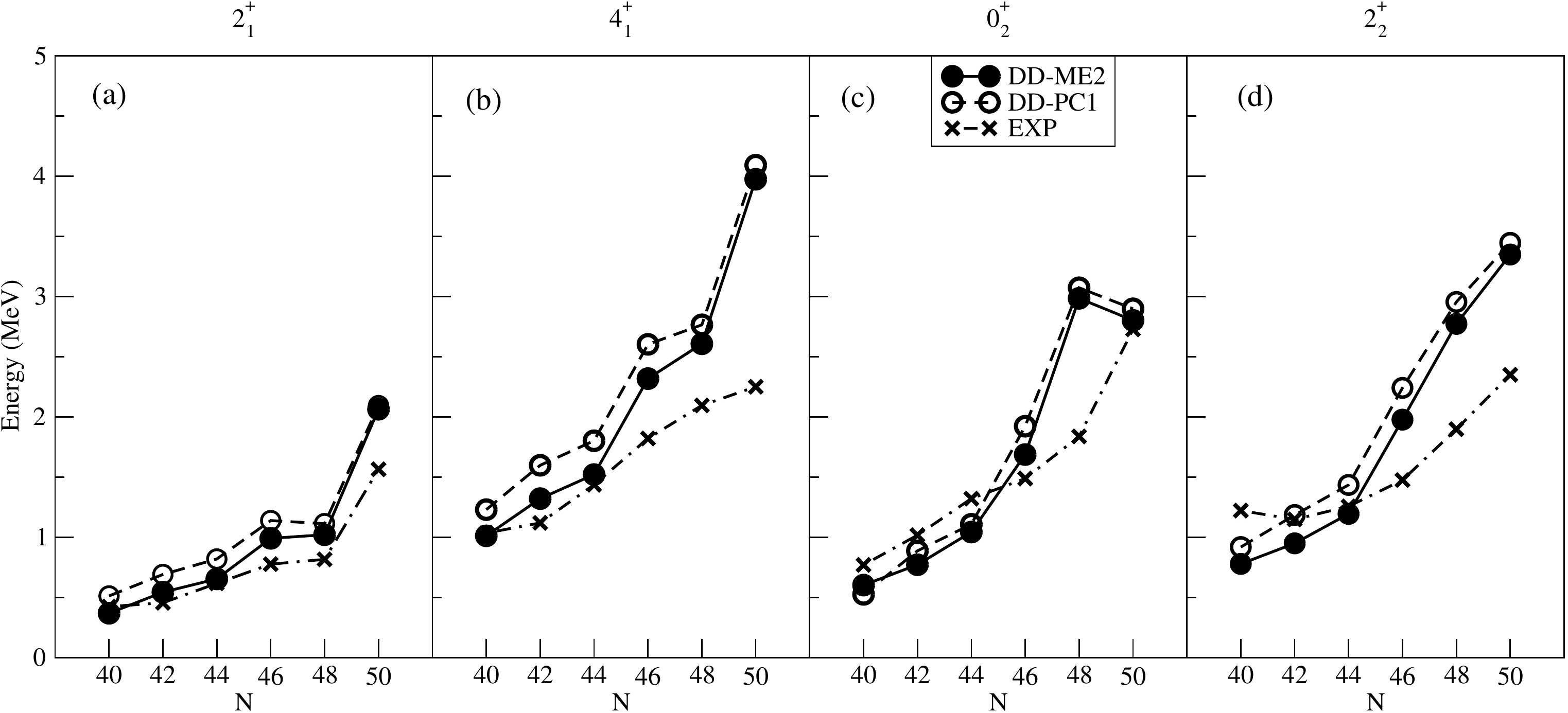}
\caption{Evolution of calculated and experimental 
low-energy spectra for the (a) $2^+_1$, (b) $4^+_1$, (c) $0^+_2$, 
and (d) $2^+_2$ states for the even-even $^{76-86}$Kr isotopes 
as functions of the neutron number $N$. The calculated 
results with both the DD-ME2 and DD-PC1 functionals 
are shown. The experimental data are taken from 
Refs.~\cite{data,rajbanshi2021}.}
\label{fig:level}
\end{center}
\end{figure*}

\begin{figure*}
\begin{center}
\begin{tabular}{cc}
\includegraphics[width=.4\linewidth]{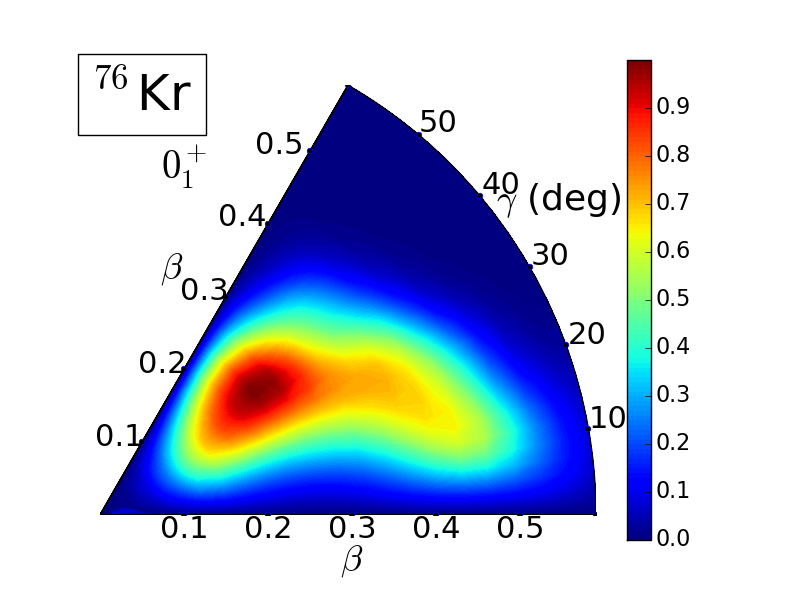} & 
\includegraphics[width=.4\linewidth]{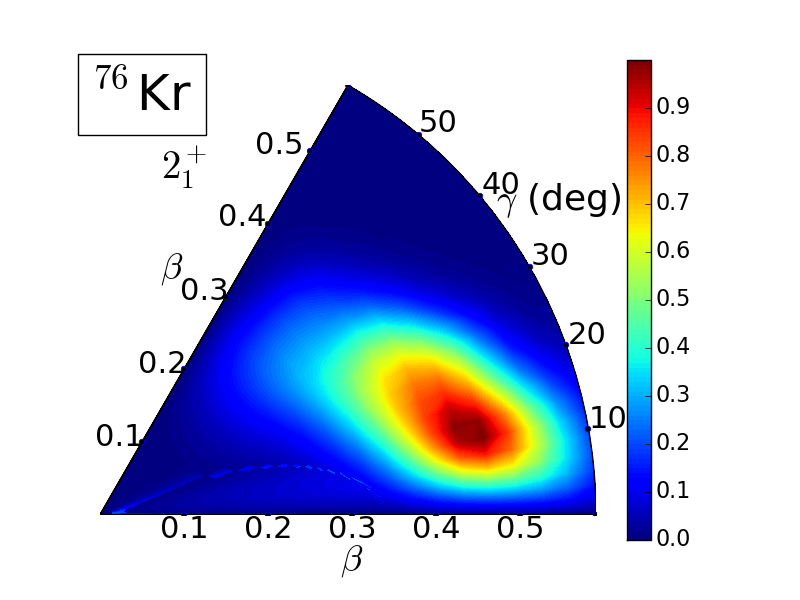} \\
\includegraphics[width=.4\linewidth]{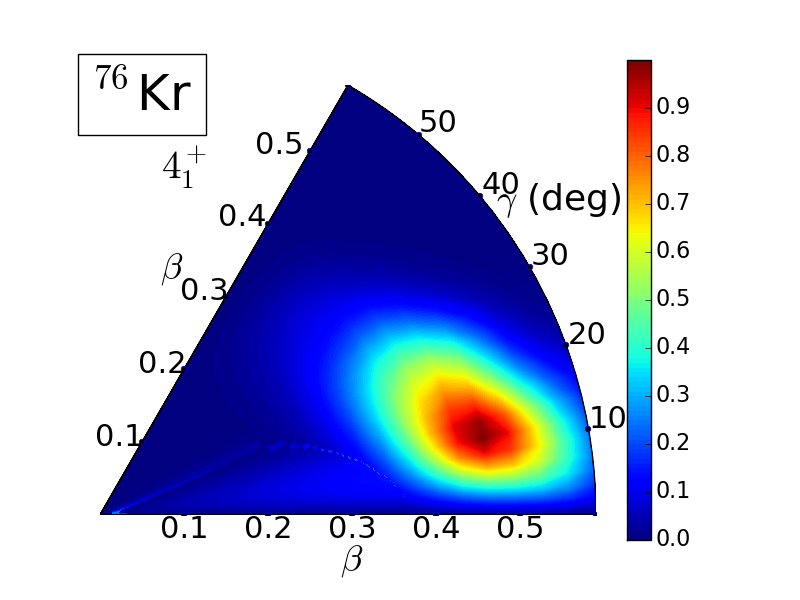} & 
\includegraphics[width=.4\linewidth]{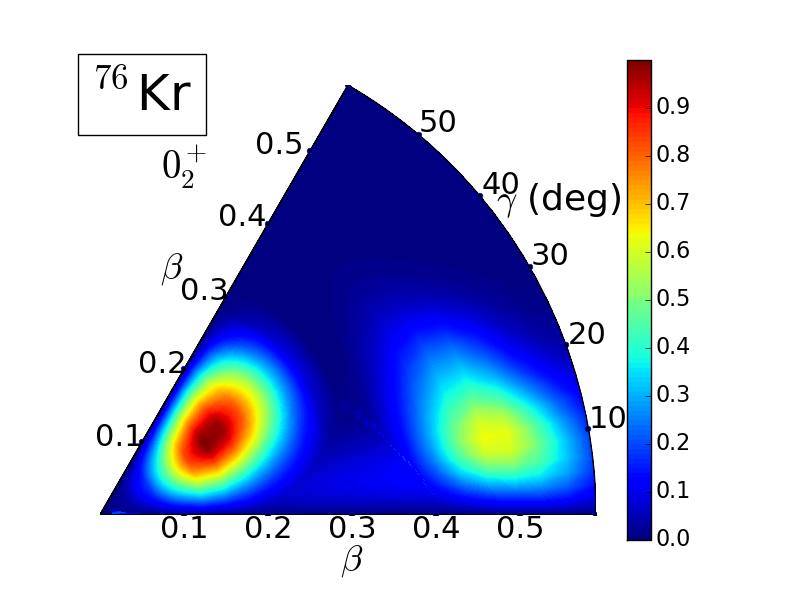} \\ 
\includegraphics[width=.4\linewidth]{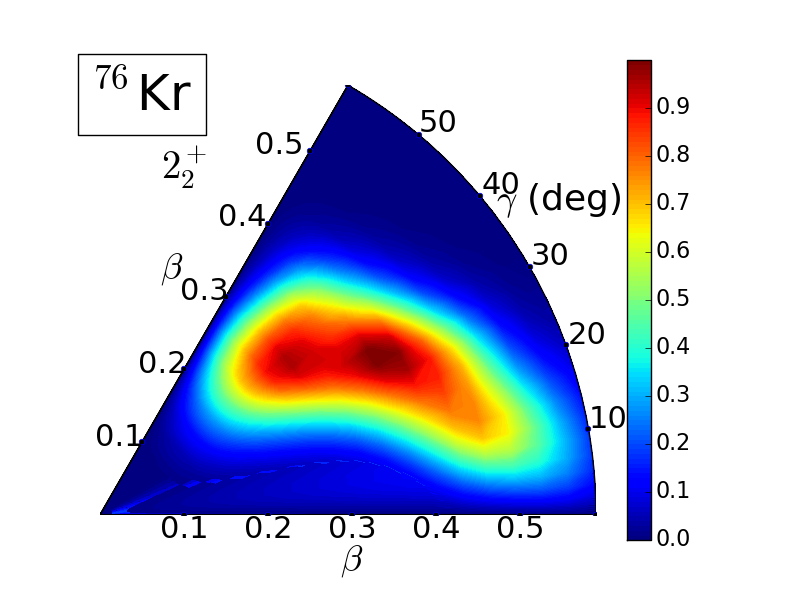} & 
\includegraphics[width=.4\linewidth]{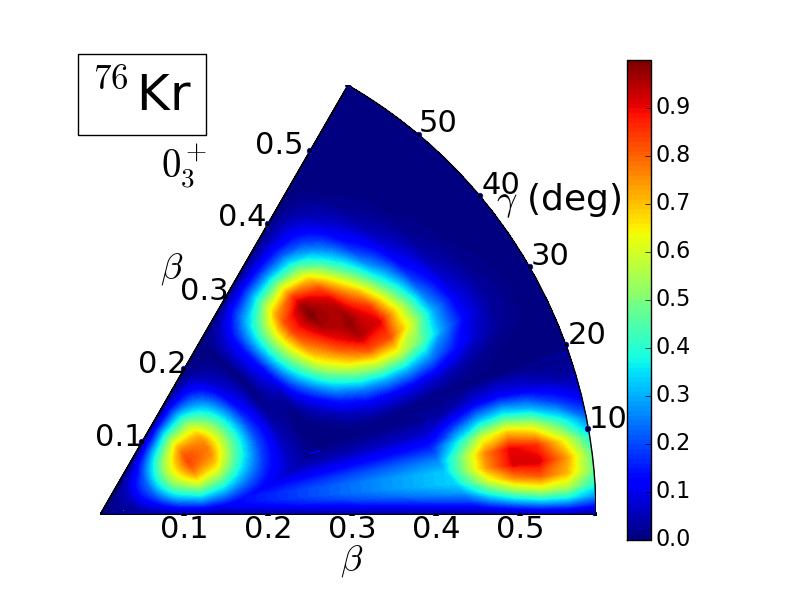} \\
\end{tabular}
\caption{
Distributions of the collective wave functions 
for the $0^+_1$, $2^+_1$, $4^+_1$, $0^+_2$, $2^+_2$, 
and $0^+_3$ states of $^{76}$Kr in the $\beta-\gamma$ plane obtained  
from the diagonalization of the QCH. The functional DD-ME2 
is used.}
\label{fig:cwf}
\end{center}
\end{figure*}

\section{Results and discussion\label{sec:results}}

\subsection{Potential energy surfaces}

Figure~\ref{fig:pes} shows contour plots of 
the triaxial quadrupole potential energy surfaces 
for the $^{76-86}$Kr nuclei defined 
in terms of the axial deformation $\beta$ and triaxiality $\gamma$. 
Only the results obtained by using the functional DD-ME2 are shown in 
the figure, because there is essentially no striking difference 
between the topology of the energy surfaces for 
the DD-ME2 and DD-PC1 EDFs. 
In Fig.~\ref{fig:pes}, for $^{76,78}$Kr, we observe that 
in addition to a (nearly) spherical global 
minimum there is also an oblate local minimum around 
$\beta\approx0.2$ on their energy surfaces. 
The spherical global minimum for $^{76}$Kr is 
separated distinctly from the oblate minimum, 
and reflects the neutron $N=40$ sub-shell closure. 
There occurs a third minimum around $\beta=0.5$ on the 
prolate axis.  
For $^{78}$Kr, the potential looks softer along 
the oblate $\gamma=60^\circ$ axis, on which the global minimum 
is identified at $\beta\approx0.05$ and the oblate local 
minimum at $\beta\approx0.2$. 
For $^{80}$Kr, the potential becomes almost completely 
flat in $\gamma$ deformation, characteristic of 
the $\gamma$-unstable O(6) symmetry of the IBM \cite{IBM}. 
Both for $^{80}$Kr and $^{82}$Kr, a weakly prolate deformed shape is 
suggested, for which the potential is still considerably 
flat in $\gamma$ direction and is soft also in the $\beta$ 
deformation. The softness implies that the fluctuations are large 
and that a significant degree of shape mixing is present 
in the vicinity of the ground state. 
Finally, an approximate harmonic oscillator potential 
with the global minimum at $\beta=0$ is obtained for $^{86}$Kr. 
This is expected, since this nucleus corresponds to 
the neutron magic number $N=50$.

\begin{figure*}[ht]
\begin{center}
\includegraphics[width=\linewidth]{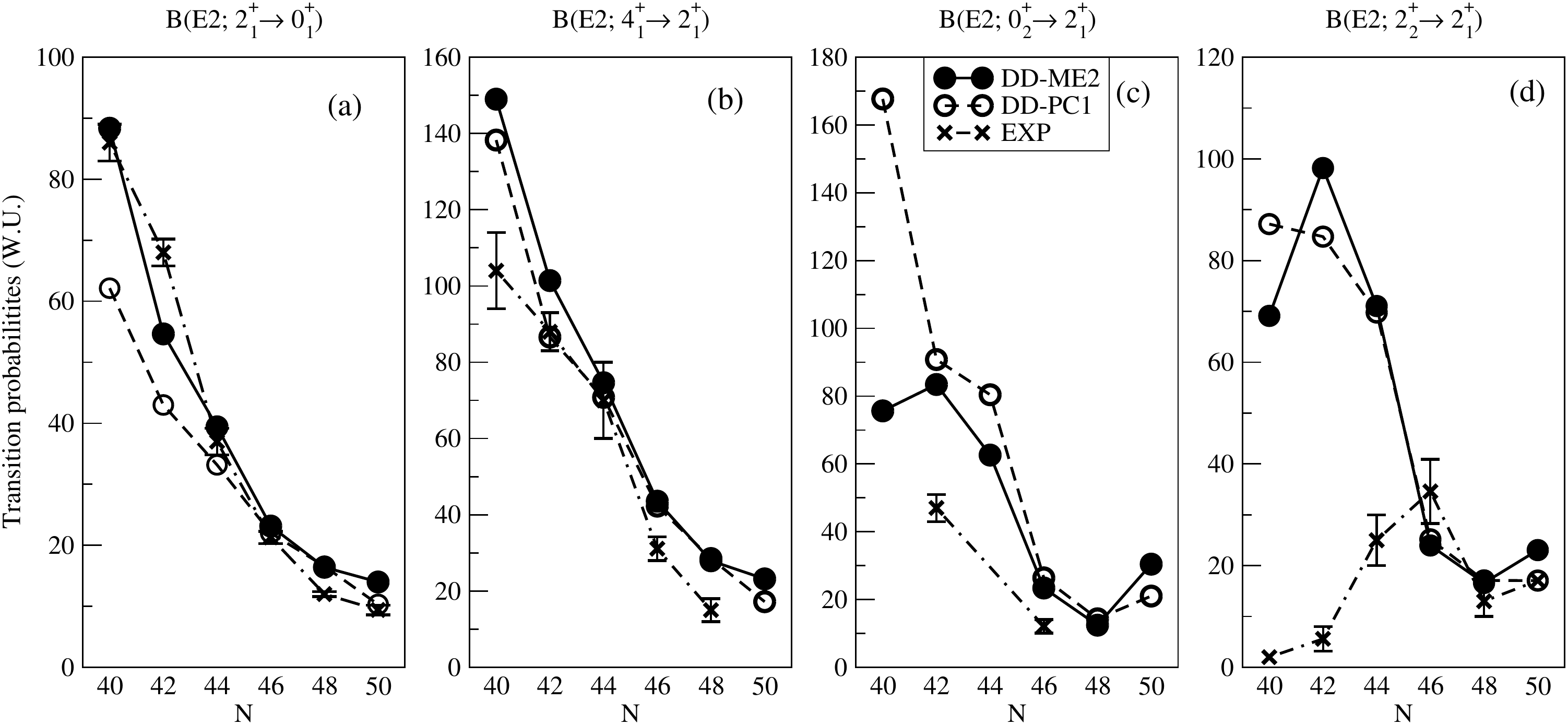}
\caption{Evolution of calculated and experimental 
$B(E2)$ strengths in Weisskopf units (W.u.) 
for the transitions 
(a) $2^+_1\to0^+_1$, (b) $4^+_1\to2^+_1$, (c) $0^+_2\to2^+_1$, 
and (d) $2^+_2\to2^+_1$ for the even-even $^{76-86}$Kr isotopes 
as functions of $N$. 
Theoretical results based on the DD-ME2 and DD-PC1 EDFs 
are shown. The experimental data are taken from Ref.~\cite{rajbanshi2021} 
for $^{82}$Kr and from Ref.~\cite{data} otherwise.}
\label{fig:e2}
\end{center}
\end{figure*}

\subsection{Systematics of low-energy spectra}

The discussion so far about the variation of the potential 
energy surface gives an approximate description of the shape 
QPT at the SCMF level, in analogy to the Landau theory of 
phase transitions. A more quantitative discussion about the QPTs 
should, therefore, involve the direct calculations of 
the spectroscopic properties that can be considered quantum 
order parameters. In the following, 
we consider overall behaviors of  
the excitation spectra and electric transition probabilities 
along the studied Kr chain.

Figure~\ref{fig:level} shows the calculated excitation 
spectra for the low-lying quadrupole collective states of the 
$^{76-86}$Kr nuclei, obtained within the RHB+QCH method. 
The results obtained with both functionals, DD-ME2 and DD-PC1 
EDFs, are shown, while there appears to be no significant qualitative or 
quantitative difference between the two functionals 
in the prediction of energy spectra. 
For comparison, the figure also includes the 
corresponding experimental data \cite{rajbanshi2021,data}. 
The RHB+QCH calculation provides a reasonable description of the  
experimental excitation energies of the $2^+_1$, $4^+_1$, 
$0^+_2$, and $2^+_2$ states for the Kr isotopes with 
the neutron numbers $40 \leq N \leq 44$. 
Both the calculated and observed $0^+_2$ excitation energy 
for the $N\approx 40$ nuclei is so low as to be 
about the same order of 
magnitude as the first excited state $2^+_1$. 
The calculation slightly underestimates the experimental 
$0^+_2$ level. The low-lying $0^+_2$ state near the neutron 
sub-shell gap $N=40$ is often considered a signature 
of shape coexistence \cite{clement2007}. 
In our model calculation, a competition between a 
nearly spherical global, an oblate and an prolate local 
minima is suggested to occur 
in the SCMF triaxial quadrupole energy 
maps for $^{76}$Kr and $^{78}$Kr (see Fig.~\ref{fig:pes}). 
For all those states shown in Fig.~\ref{fig:level}, 
the predicted excitation energies turn to increase abruptly 
from $N=44$ to 46, and overestimate the data. 
As we approach the neutron major shell gap $N=50$, both the 
calculated and experimental energy levels become higher with $N$. 
The energy levels of the non-yrast states 
$0^+_2$ and $2^+_2$ are here predicted to be particularly 
higher than the experimental ones for $N=48$ and 50. 
Since the collective Hamiltonian gives purely collective states, 
the description of those nuclei close to the magic numbers 
becomes worse, in which nuclei single-particle excitations 
play a more relevant role. 

The collective wave function, 
resulting from the diagonalization of the QCH, 
provides an insight into 
the nature of a given low-lying state. Of particular interest 
is $^{76}$Kr, for which three competing mean-field minima 
occur in the potential energy surface 
(see Fig.~\ref{fig:pes}). 
Figure~\ref{fig:cwf} shows contour plots of the collective 
wave functions in the $\beta-\gamma$ 
deformation plane corresponding to the $0^+_1$, $2^+_1$, $4^+_1$, 
$0^+_2$, $2^+_2$, and $0^+_3$ states. 
The $0^+_1$ wave function is spread over an area 
from the oblate to prolate sides, and exhibits a peak 
that is close to the oblate axis. The $2^+_1$ wave function 
is, on the other hand, more sharply peaked on the prolate 
side with the deformation $\beta\approx0.5$, around which the 
prolate local minimum occurs in the potential energy surface. 
The same is true for the $4^+_1$ state and 
those with higher spin, $I>4$, of the ground-state band. 
This result implies a transition from the nearly oblate 
to prolate configurations at low spin within the ground-state band, 
and thus the higher-spin members of the band are supposed to 
be made mainly of the strongly deformed prolate configurations. 
This finding is consistent with the conclusion drawn 
from the earlier cranking RHB calculation of 
Ref.~\cite{afanasjev2005}. 
The distribution of the $0^+_2$ wave function in the 
$\beta-\gamma$ surface indicates 
a distinct coexistence between the oblate and prolate 
shapes. The $2^+_2$ state is suggested to be made largely 
of the triaxial configurations around $\gamma=30^\circ$. 
One observes essentially three peaks in the $0^+_3$ 
wave function distribution, which are associated with 
the weakly triaxially deformed, and the nearly oblate and 
prolate deformed configurations. 

In addition, it is meaningful to study the sensitivity 
of the predicted excitation spectra to the 
pairing strength. A global study of the separable pairing 
force within the covariant density functional framework 
in Ref.~\cite{teeti2021} indicated that, in order to account 
for the empirical odd-even mass staggering, the strength 
of the separable pairing force needs to be modified 
so that it is scaled by particle-number 
dependent factors. We have then carried out the 
RHB+QCH calculation in which both the proton 
and neutron pairing strengths are scaled with the factors 
introduced in Eqs.~(13--17) of Ref.~\cite{teeti2021}.  
For the $^{76}$Kr and $^{82}$Kr isotopes, for example, 
this modification gives rise to an increase of the 
pairing strengths by approximately 15 \%, if the parameters 
listed in Table I in that reference are adopted. 
For both of these nuclei, the RHB+QCH calculation employing 
the increased pairing strengths gives excitation spectra 
for all the states that are systematically larger than those 
obtained when the original pairing strength $V_0=728$ MeV fm$^3$ 
is employed. 
Thus, in this particular case, the use of the  
pairing strength that is increased according to the prescription 
of Ref.~\cite{teeti2021} does not appear to improve the 
description of the excitation energies. 

\begin{figure}
\begin{center}
\includegraphics[width=\linewidth]{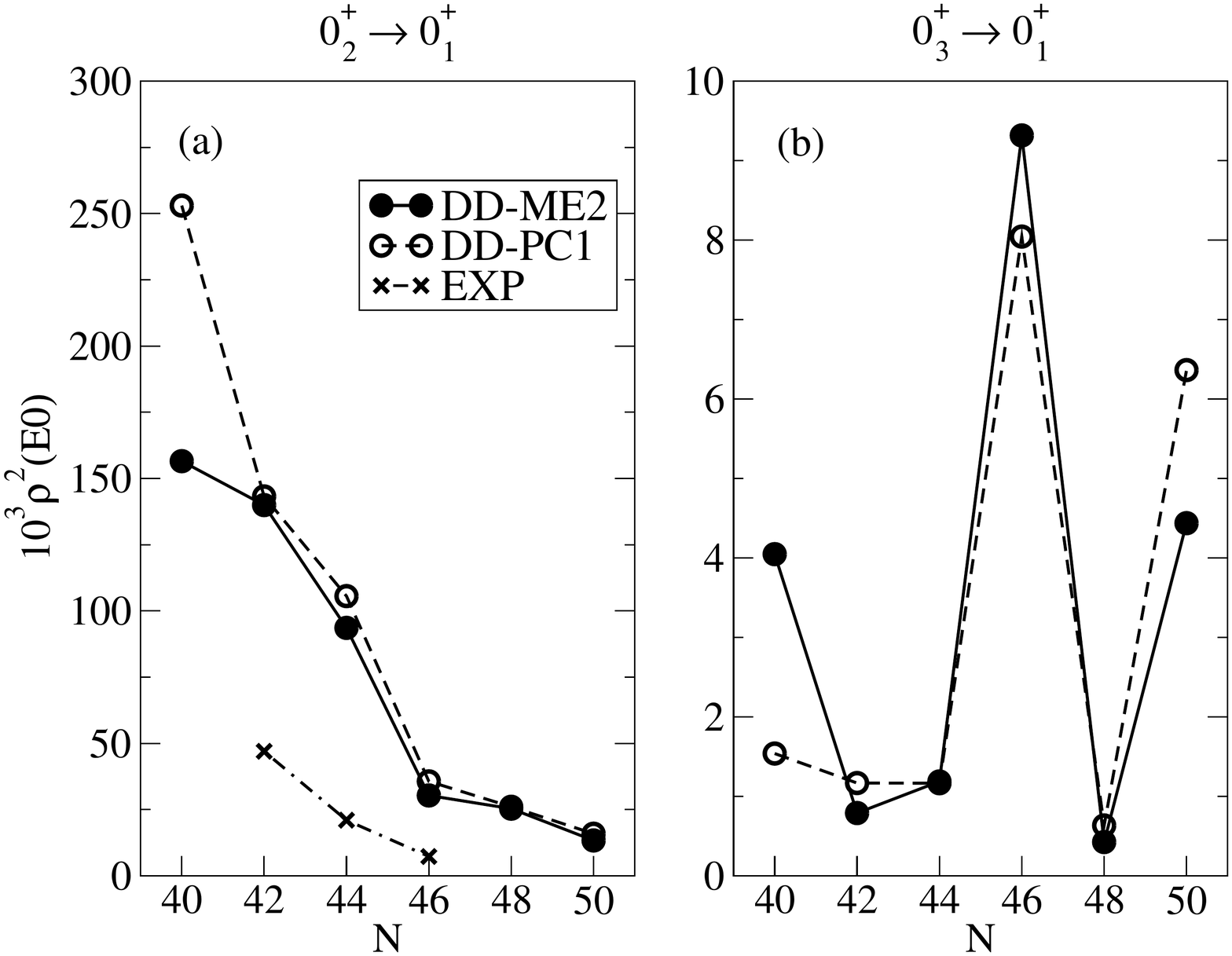}
\caption{Calculated and experimental 
$\rho^2(E0;0^+_2\to0^+_1)$ and $\rho^2(E0;0^+_3\to0^+_1)$ 
values for the $^{76-86}$Kr nuclei. The experimental 
values are taken from Ref.~\cite{kibedi2005}.}
\label{fig:e0}
\end{center}
\end{figure}

\begin{figure*}
\begin{center}
\includegraphics[width=\linewidth]{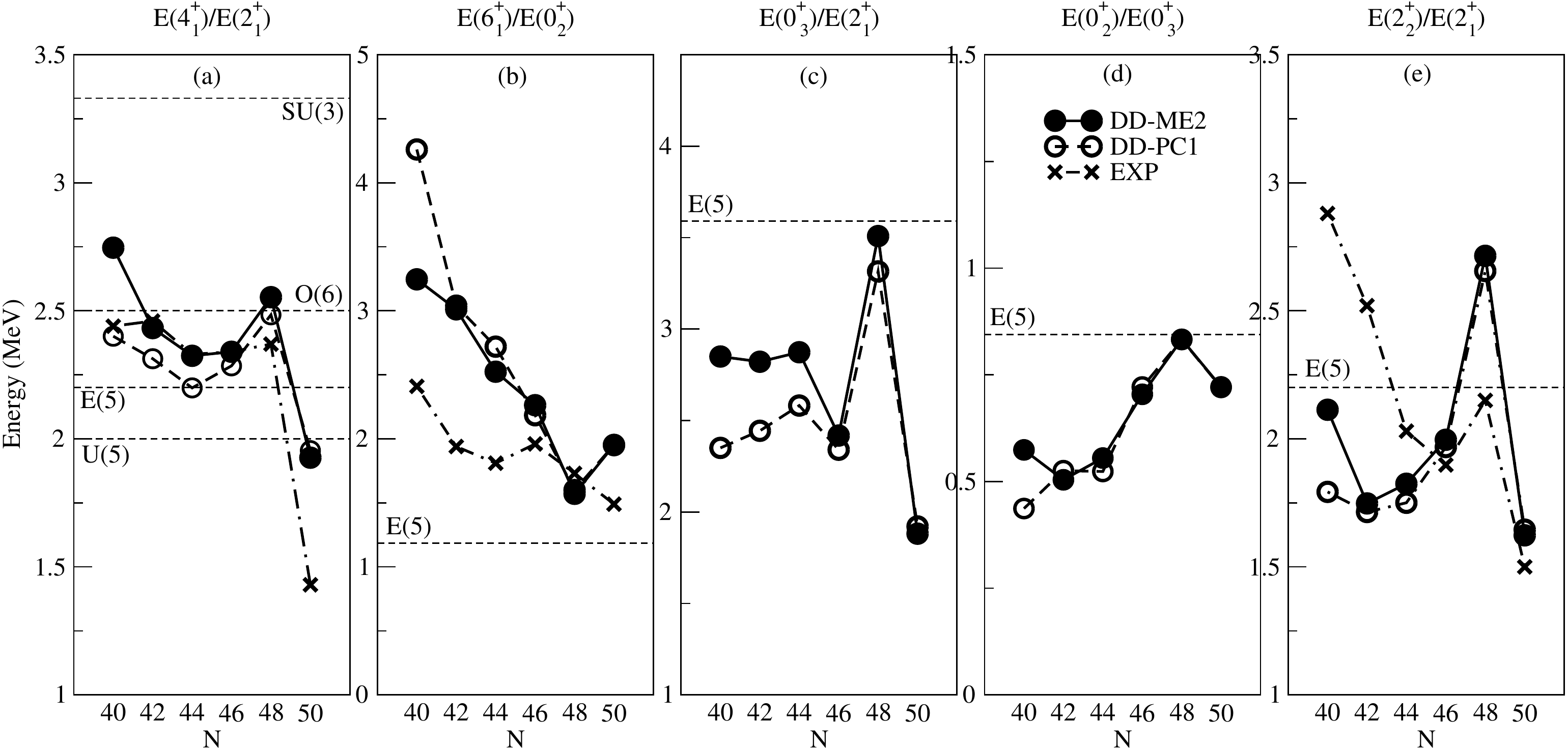}
\caption{Evolution of calculated and experimental 
energy ratios (a) $E(4^+_1)/E(2^+_1)$, 
(b) $E(6^+_1)/E(0^+_2)$, (c) $E(0^+_3)/E(2^+_1)$, 
(d) $E(0^+_2)/E(0^+_3)$, and (e) $E(2^+_2)/E(2^+_1)$ 
for the even-even $^{76-86}$Kr isotopes as functions of $N$. 
The calculated results with both the DD-ME2 and DD-PC1 functionals 
are shown, in comparison to the limits of the E(5) symmetry and 
three dynamical symmetries of the IBM.}
\label{fig:eratio}
\end{center}
\end{figure*}

\subsection{Systematics of $B(E2)$ transition rates\label{sec:e2}}

In Fig.~\ref{fig:e2} we show the results for the $B(E2)$ 
rates for the electric quadrupole transitions between 
the low-lying states, 
i.e., $B(E2;2^+_1\to0^+_1)$, $B(E2;4^+_1\to2^+_1)$, 
$B(E2;0^+_2\to2^+_1)$, and $B(E2;2^+_2\to2^+_1)$. 
The RHB+QCH calculation reproduces the 
experimental data for 
the $B(E2;2^+_1\to0^+_1)$ and $B(E2;4^+_1\to2^+_1)$ 
rates fairly well. 
The decreasing pattern of the $B(E2;2^+_1\to0^+_1)$ 
and $B(E2;4^+_1\to2^+_1)$ values suggests the weakening of 
the quadrupole collectivity towards the neutron magic number $N=50$. 
For those nuclei with $40 \leq N \leq 44$, 
the present calculation suggests much stronger 
interband $E2$ transitions $0^+_2\to2^+_1$ and $2^+_2\to2^+_1$  
than the experimental data. 
This result further confirms that a high degree 
of shape mixing is present near the ground state of 
these nuclei, especially $^{76}$Kr and $^{78}$Kr: 
the corresponding 
SCMF potential energy surfaces 
are notably soft in both the $\beta$ and 
$\gamma$ deformations, and indicate the coexistence 
of the three minima (see Fig.~\ref{fig:pes}). 
While the energy levels of the $0^+_2$ 
[Fig.~\ref{fig:level}(c)] and $2^+_2$ [Fig.~\ref{fig:level}(d)] 
states are reasonably described, 
the interband $E2$ transitions appear to be rather 
sensitive to the relevant wave functions. 
One can see in Fig.~\ref{fig:cwf}, for instance, 
a substantial overlap between the $2^+_1$ 
and $2^+_2$ collective wave functions in $^{76}$Kr, 
which can result in the too enhanced $2^+_2\to2^+_1$ 
transition as compared to the experimental value. 
For the $N\geq 46$ nuclei, on the other hand, 
we observe that the calculated  
$B(E2;0^+_2\to2^+_1)$ and $B(E2;2^+_2\to2^+_1)$ values 
[Figs.~\ref{fig:e2}(c) and \ref{fig:e2}(d)] rapidly 
decrease from $N=44$ to 46, and are consistent with 
the experimental values. 
Note, however, that the decrease of the interband 
$B(E2)$ rates is also considered a consequence of the fact that 
the quadrupole collectivity becomes weaker as the neutron 
major shell closure $N=50$ is approached. 

The calculated results for the $B(E2)$ transition 
rates based on the two EDFs, DD-ME2 and DD-PC1, are 
basically similar to each other 
both at the qualitative and quantitative levels. 
A notable difference, however, arises in the prediction 
of the $B(E2;0^+_2\to2^+_1)$ values at $N=40$, for which the DD-PC1 EDF 
leads to about twice as large a value as the DD-ME2 EDF.

\subsection{$E0$ transitions\label{sec:e0}}

Let us consider the monopole transition properties of the 
studied Kr nuclei. 
Figure~\ref{fig:e0} shows the calculated $\rho^2(E0)$ values 
for the $E0$ transitions $0^+_2\to0^+_1$ and $0^+_3\to0^+_1$. 
The $\rho^2(E0;0^+_2\to0^+_1)$ values are here calculated to be 
considerably larger than the experimental values 
\cite{kibedi2005}, while the observed decreasing pattern 
from $N=42$ to 46 is reproduced by our model calculation. 
The large $E0$ transition strengths, especially for 
the Kr nuclei with $N=40-44$, corroborates the strong shape 
mixing between the wave functions of the low-lying $0^+$ states. 
See in Fig.~\ref{fig:cwf} a significant overlap between the 
$0^+_1$ and $0^+_2$ collective functions for $^{76}$Kr. 
It is worth noticing that there appears a sudden decrease 
of the calculated $\rho^2(E0;0^+_2\to0^+_1)$ value from $N=44$ to 46, 
implying a rapid nuclear structure change. 
As seen in Fig.~\ref{fig:e0}(b), 
in the present calculation the $\rho^2(E0;0^+_3\to0^+_1)$ value 
is by more than two orders of magnitude smaller than the 
$\rho^2(E0;0^+_2\to0^+_1)$ one, and hence no large overlap 
between the $0^+_3$ and the $0^+_1$ ground state is expected 
to be present. 
Similarly to the $B(E2)$ results, the most notable difference 
between the theoretical $\rho^2(E0;0^+_3\to0^+_1)$ 
values obtained from the two EDFs appears at $N=40$. 

\begin{figure*}
\begin{center}
\includegraphics[width=\linewidth]{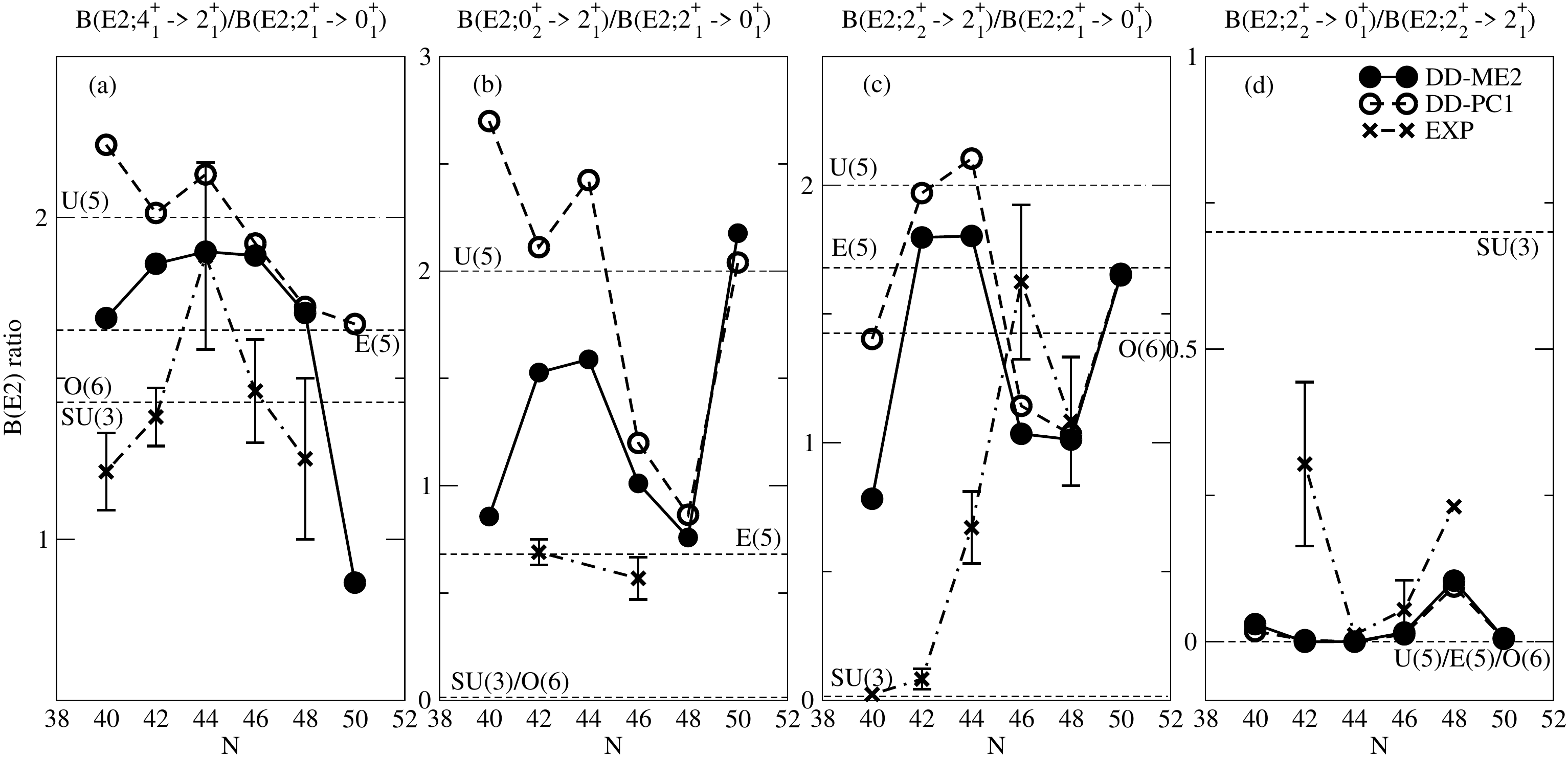}
\caption{Evolution of calculated and experimental 
ratios of the $B(E2)$ transition rates, 
(a) $R_1=B(E2; 4^+_1\to2^+_1)/B(E2; 2^+_1\to0^+_1)$, 
(b) $R_2=B(E2; 0^+_2\to2^+_1)/B(E2; 2^+_1\to0^+_1)$, 
(c) $R_3=B(E2; 2^+_2\to2^+_1)/B(E2; 2^+_1\to0^+_1)$, and 
(d) $R_4=B(E2; 2^+_2\to0^+_1)/B(E2; 2^+_2\to2^+_1)$, 
for the $^{76-86}$Kr isotopes. 
The values predicted by the dynamical symmetries of the 
IBM (U(5), SU(3), and O(6)) and the E(5) CPS are also indicated.
The experimental data are
taken from Ref.~\cite{data,rajbanshi2021}
}
\label{fig:e2ratio}
\end{center}
\end{figure*}

\begin{figure}[ht]
\begin{center}
\includegraphics[width=.8\linewidth]{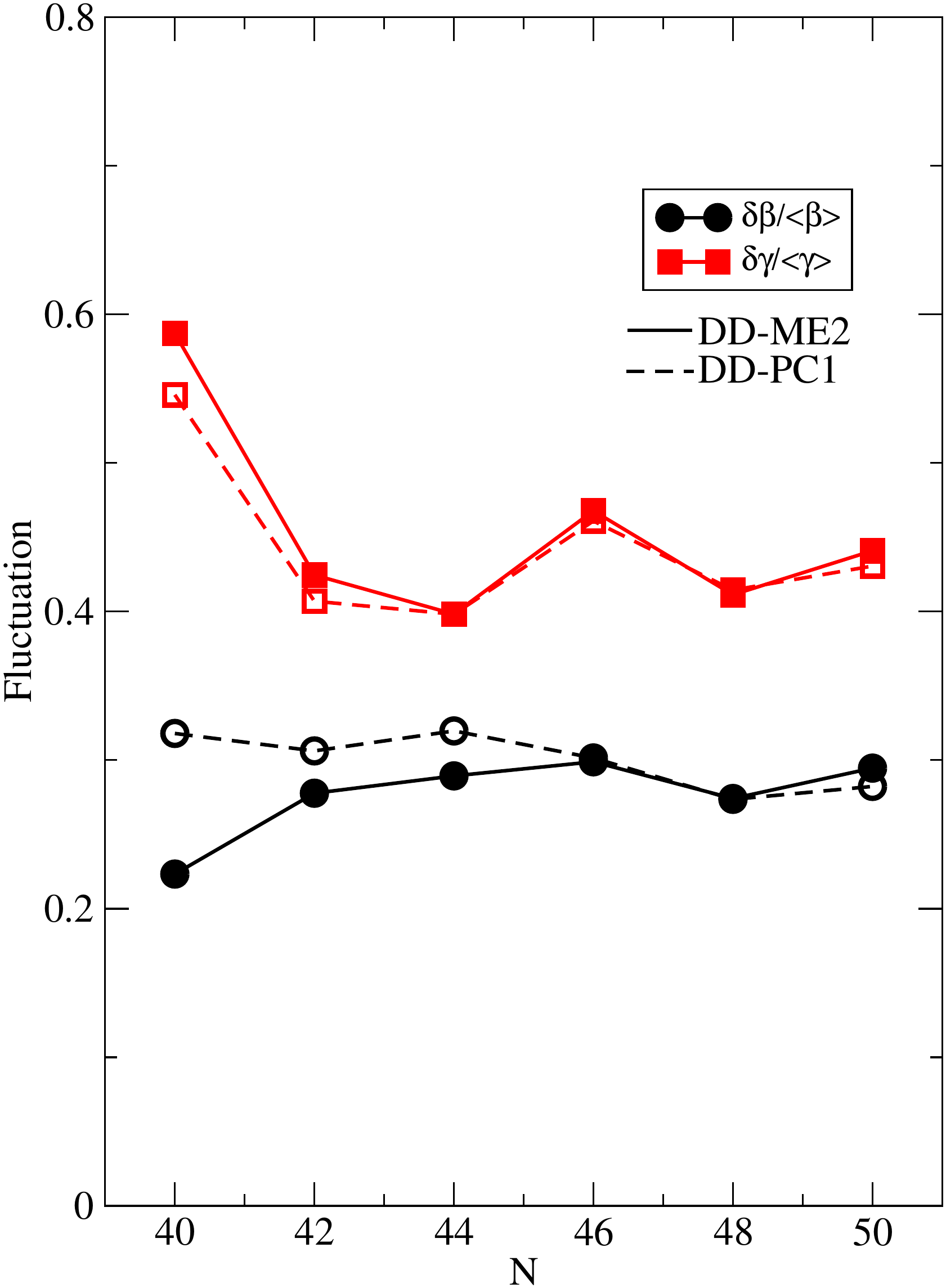}
\caption{Fluctuations of the $\beta$ and $\gamma$ 
deformations for $^{76-86}$Kr, computed by using the QCH 
wave function of the $0^+_1$ ground state for each nucleus 
based on the DD-ME2 and DD-PC1 EDFs.}
\label{fig:fluc}
\end{center}
\end{figure}

\subsection{Signatures of shape phase transitions}

We have seen in the previous sections 
that the RHB+QCH calculation provides a 
fairly reasonable description of the observed low-energy spectra, 
$B(E2)$, and $\rho^2(E0)$ values. 
Let us now turn to analyze several spectroscopic properties 
that can serve as a more distinct signature of the QPT, 
especially, in comparison with various 
symmetry limits of the IBM and E(5). 
Such an analysis also presents 
a sensitive test of the employed model.

\subsubsection{Energy ratios}

Figure~\ref{fig:eratio} shows ratios of the calculated 
excitation energies of low-lying states for the considered 
$^{76-86}$Kr nuclei. 
The ratio $R_{4/2}\equiv E(4^+_1)/E(2^+_1)$ is a typical  
indicator that distinguishes among 
various geometrical limits for the nuclear shapes. 
In Fig.~\ref{fig:eratio}(a) the calculated ratio $R_{4/2}$ exhibits a 
weak parabolic dependence on $N$, with a minimum value at $N=44$. 
In most of the nuclei, the calculated $R_{4/2}$ ratio is 
approximately in between the E(5) limit, $R_{4/2}=2.20$, 
and the $\gamma$-unstable O(6) limit of the IBM, 
$R_{4/2}=2.5$ \cite{IBM}. 
For both EDFs, the $R_{4/2}$ values for 
$^{80}$Kr and $^{82}$Kr in the RHB+QCH calculation 
appear to be close to the E(5) limit $R_{4/2}=2.20$.

In Fig.~\ref{fig:eratio}(b) the observed energy ratio 
$R_{6/0}\equiv E(6^+_1)/E(0^+_2)$ 
gradually decreases with $N$, approaching the E(5) limit 1.19. 
The calculated values with both EDFs show a similar, but 
a more rapid change with $N$. 
This trend also conforms to the shape evolution into 
$\gamma$-soft deformation in the studied Kr nuclei.

In Fig.~\ref{fig:eratio}(c), the energy ratio 
$R_{0/2}\equiv E(0^+_3)/E(2^+_1)$ is shown. 
For most of the nuclei the RHB+QCH calculation provides 
smaller values than the E(5) limit $R_{0/2}=3.59$. 
The calculation with the DD-ME2 
EDF gives larger $R_{0/2}$ ratios for $40 \leq N \leq 44$. 
The calculation, however, suggests 
an irregular behavior of $R_{0/2}$ from $N=46$ to 50. 
This is mostly because 
both the $2^+_1$ and $0^+_3$ excitation energies are here 
overestimated, especially as the neutron number increases towards the 
$N=50$ neutron magic number (see Fig.~\ref{fig:level}(a)). 

Figure~\ref{fig:eratio}(d) shows the ratio 
$R_{0/0}\equiv E(0^+_2)/E(0^+_3)$, which provides information about 
relative locations of the first and second excited $0^+$ levels. 
The calculated ratio $R_{0/0}=0.704$ (or 0.721) 
in the case of the DD-ME2 (or DD-PC1) EDF for $^{82}$Kr is 
in a fair agreement with the E(5) limit, $R_{0/0}=0.84$, 
and with the experimental data, 0.9 \cite{rajbanshi2021}. 
From Fig.~\ref{fig:eratio}(d) 
the calculated $R_{0/0}$ values for $^{84}$Kr and $^{86}$Kr 
are also close to the E(5) limit. 
However, this is simply because 
the $0^+_{2,3}$ excitation energies for 
these nearly spherical nuclei are not satisfactorily described 
by the collective Hamiltonian, and should not be considered 
a signature of the E(5) symmetry.

The energy ratio $R_{2/2}\equiv E(2^+_2)/E(2^+_1)$ 
indicates the location of the $\gamma$-vibrational band, 
with the $2^+_2$ state being the bandhead, 
relative to the ground-state $K=0^+_1$ band. 
As shown in Fig.~\ref{fig:eratio}(e), the calculated $R_{2/2}$ ratios
underestimate the experimental data for the region $40\leq N \leq 44$, 
while the isotopic dependence looks more or less similar 
between the theory and experiment. 
The predicted value $R_{2/2}=2.00$ (or 1.97) 
with the DD-ME2 (or DD-PC1) 
EDF is close to both the experimental data, 1.90, 
and the E(5) limit, 2.20.

\subsubsection{$B(E2)$ ratios}

Next, we consider the ratios of the calculated 
$B(E2)$ transition rates: 
\begin{align}
&R_1=B(E2; 4^+_1\to2^+_1)/B(E2; 2^+_1\to0^+_1), \\
&R_2=B(E2; 0^+_2\to2^+_1)/B(E2; 2^+_1\to0^+_1), \\
&R_3=B(E2; 2^+_2\to2^+_1)/B(E2; 2^+_1\to0^+_1), \\
&R_4=B(E2; 2^+_2\to0^+_1)/B(E2; 2^+_2\to2^+_1).
\end{align}
The corresponding results are shown in Fig.~\ref{fig:e2ratio}, 
and are compared with the U(5), SU(3), and O(6) 
limits of the IBM, and the E(5) \cite{iachello2000} limit. 
Note that the three IBM limits shown in the figure 
are obtained in the large-boson-number limit \cite{IBM}.

As seen in Fig.~\ref{fig:e2ratio}(a), 
the $R_1$ ratios obtained here are approximately in between the 
U(5) vibrational limit, $R_1=2.0$, and the E(5) limit, $R_1=1.68$.
Both functionals lead to systematically larger $R_1$ values 
than the experimental ones, 
except for $^{80}$Kr in the case of the DD-ME2 result. 
The calculation with DD-ME2 EDF 
generally produces the $R_1$ values closer 
to the data than with the DD-PC1. 
The systematic trend of the $R_1$ ratio with $N$ in the DD-ME2 
case also looks more or less similar to the experimental one, 
i.e., an inverse parabolic dependence on $N$ centered around $N=44$.

The calculated $R_2$ ratios are shown in Fig.~\ref{fig:e2ratio}(b). 
The experimental data are available only 
for the $^{78}$Kr and $^{82}$Kr nuclei.  
As anticipated by the strong $0^+_2\to2^+_1$ transitions 
[cf. Fig.~\ref{fig:e2}(c)], 
the present RHB+QCH calculation overestimates the experimental 
$R_2$ ratios by several factors for these nuclei. 
The ratio $R_{2}$ also seems to be quite sensitive 
to the choice of the EDFs for $40 \leq N \leq 44$. 
In particular, 
the calculation with the DD-ME2 EDF generally yields 
a smaller $R_2$ ratio, hence closer to the experimental value, 
than with the DD-PC1. 
Both EDFs produce rather small $R_{2}$ values for $^{82}$Kr 
and $^{84}$Kr. 
Especially for the former nucleus, our calculation gives 
$R_2=1.01$ (or 1.20) with the DD-ME2 (or DD-PC1) functional, 
while the experimental and E(5) values are 
$R_2=0.57\pm0.10$ and $R_2=0.68$, respectively. 

Figure~\ref{fig:e2ratio}(c) shows the results for the $R_3$ ratio. 
Experimentally,  
this quantity appears to reflect the structural 
evolution from the SU(3) rotational limit at $N=40$ and 42 
to the $\gamma$-unstable O(6) or E(5) limit at $N=46$. 
Concerning the $^{76-80}$Kr nuclei, the calculated $R_3$ ratios 
are much larger than the data, and are 
also quite far from the SU(3) limit $R_3=0$. 
The large finite $R_3$ ratios for these nuclei 
further confirm the enhanced shape mixing in the present theoretical framework, 
as is shown in Fig.~\ref{fig:pes} that the SCMF potential 
energy surfaces have coexisting mean-field minima for $^{76,78}$Kr
and are totally flat in the $\gamma$ direction for $^{80}$Kr. 
As a consequence, the overlap between the resultant wave 
functions for the low-spin states is supposed to be large, 
leading to the unexpectedly strong $2^+_2\to2^+_1$ $E2$ transitions 
for the $^{76-80}$Kr nuclei. 
Probably the low-lying structures of these nuclei are 
so complicated that the simple QCH approach combined 
with a particular choice of the underlying EDF and pairing 
interaction may not necessarily 
give a reasonable agreement with the empirical data. 
For those nuclei with $N \geq 46$, however, 
the predicted $R_3$ ratios agree rather well with the 
experimental data. Note that the value for $^{82}$Kr has 
a large error bar.

The $R_4$ ratio particularly distinguishes the 
deformed rotor limit SU(3) from the O(6) and U(5) ones. 
As seen in Fig.~\ref{fig:eratio}(d), 
the present calculation provides, for most of the Kr nuclei, 
nearly vanishing values of $R_4$. 
Our results agree with the data for $^{80,82,84}$Kr, but 
not for $^{78}$Kr. 
For the latter nucleus, a large finite value $R_4=0.30\pm0.14$ 
has been observed experimentally \cite{data}. 
The vanishing $R_4$ ratio obtained by our calculation 
for the $N \leq 44$ nuclei reflects that 
the corresponding $B(E2;2^+_2\to2^+_1)$ 
rates are calculated to be considerably large due to the strong 
configuration mixing [see Fig.~\ref{fig:e2}(d)].

\subsubsection{Fluctuations in shape variables}

As yet another signature of the QPT in the Kr isotopes, 
we consider the fluctuations for the 
$\beta$ and $\gamma$ deformations, defined respectively as 
$\delta\beta/\braket{\beta}$ and 
$\delta\gamma/\braket{\gamma}$. 
Here, 
\begin{align}
&\delta\beta=\frac{\sqrt{\braket{\beta^4}-\braket{\beta^2}^2}}{2\braket{\beta}} \; ,\\
&\delta\gamma = 
\left(
\sqrt{\frac{\braket{\beta^6\cos^2{3\gamma}}}{\braket{\beta^6}}}-\sqrt{\frac{\braket{\beta^3\cos{3\gamma}}^2}{\braket{\beta^4}\braket{\beta^2}}}
\right) \nonumber \\
&\quad
\times
\left(
{3\sin{3\braket{\gamma}}}
\right)^{-1} \; ,
\end{align}
stand for the deviation, and  
\begin{align}
&\braket{\beta}=\sqrt{\braket{\beta^2}} \\
&\braket{\gamma}=\arccos
{\left(
\braket{\beta^3\cos{3\gamma}}/\sqrt{\braket{\beta^4}\braket{\beta^2}}
\right)/3}
\end{align}
are the average values of the $\beta$ and $\gamma$ 
deformations, respectively. The above quantities are 
computed by using the wave function for the $0^+_1$ ground 
state. The fluctuations in the deformation variables 
have also been considered in previous EDF-based calculations 
for the studies, e.g., of the spherical to $\gamma$-soft shape 
transitions in Ba and Xe nuclei in the mass $A\approx130$ region 
\cite{li2010}, and of the quadrupole-octupole shape phase 
transitions in a wider mass region \cite{nomura2021qoch}. 
A discontinuity of the fluctuations when plotted as functions 
of the nucleon number is considered a signature 
of the QPT. 

Figure~\ref{fig:fluc} shows the corresponding results 
obtained from both the DD-ME2 and DD-PC1 EDFs. 
We see that, regardless of the choice of the EDF, 
the fluctuation in the 
$\gamma$ deformation $\delta\gamma/\braket{\gamma}$ 
exhibits a notable kink at $N=46$, 
signaling the occurrence of the QPT. 
There also appears a significant decrease of the 
$\gamma$ fluctuation from $N=40$ to 42, 
indicating the effect of the neutron sub-shell closure $N=40$. 
According to the SCMF results 
presented in Fig.~\ref{fig:pes}, the potential energy surface for 
$^{80}$Kr is almost completely flat in the $\gamma$ deformation, 
while the global prolate minimum at $\beta\approx0.15$ appears 
for $^{82}$Kr. 
In Fig.~\ref{fig:fluc} the fluctuation in the axial deformation 
$\beta$, $\delta\beta/\braket{\beta}$, 
shows a minor kink at $N=46$, when the DD-ME2 EDF is 
chosen. In general, however, the fluctuation in the axial 
deformation $\delta\beta/\braket{\beta}$ shows only a gradual 
variation with $N$, and is not considered 
as distinct a signature as the one for the triaxial 
deformation $\delta\gamma/\braket{\gamma}$.

\begin{figure*}[ht]
\begin{center}
\includegraphics[width=\linewidth]{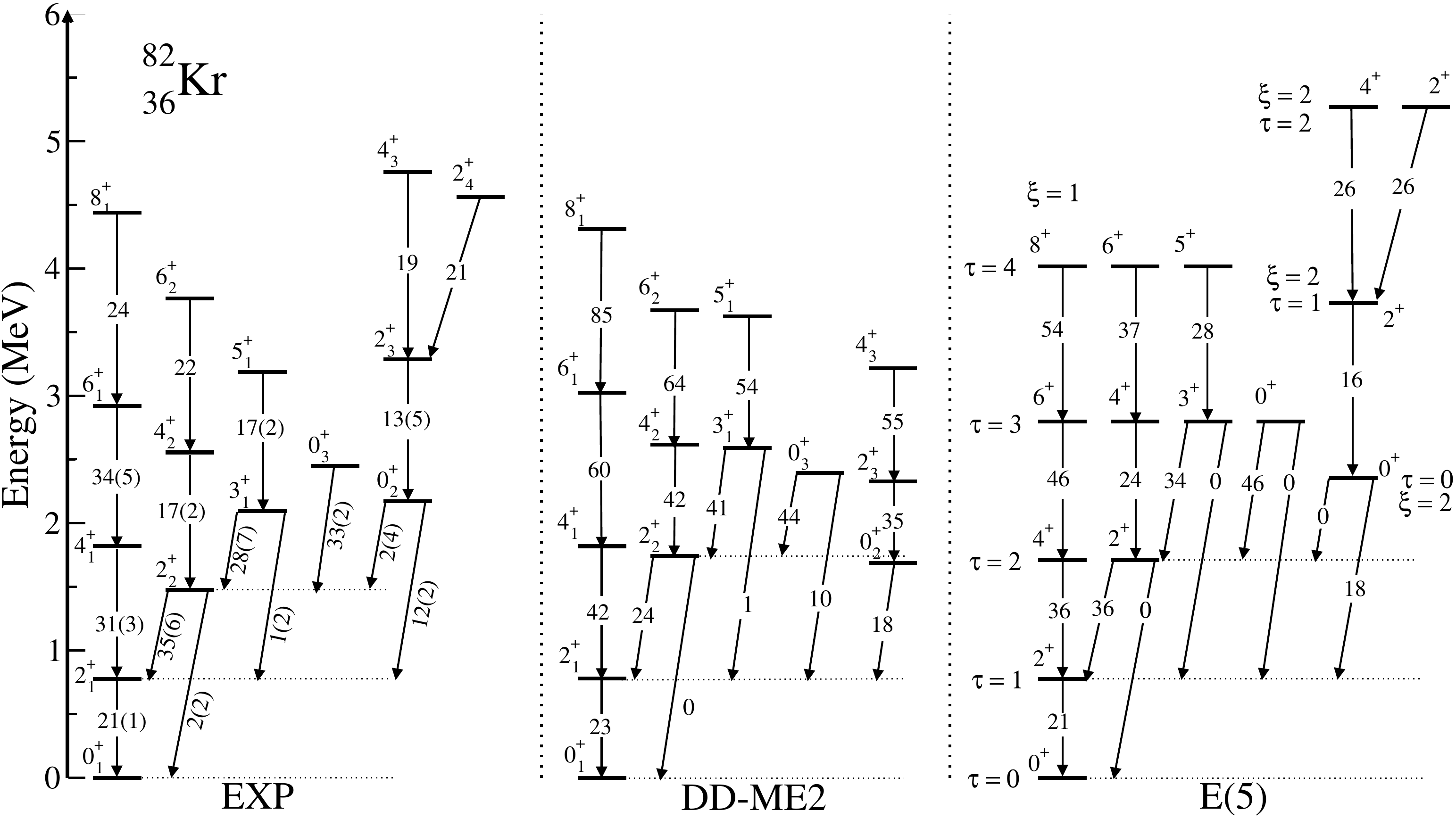}
\caption{Comparison of experimental, calculated, and E(5) energy 
spectra and $B(E2)$ transition strengths (in W.u.) for the 
$^{82}$Kr nucleus. The functional DD-ME2 is used 
for the theoretical spectrum. Note that, in the QCH calculation, 
the IB moments of inertia are increased by 40 \%. 
The experimental data are 
taken from Ref.~\cite{rajbanshi2021}. 
In the E(5) spectrum, quantum 
numbers labelling states $\xi$ and $\tau$ are shown, 
and the excitation energy of the 
state $2^+_{\xi=1,\tau=1}$ and the $B(E2;2^+_{1,1}\to0^+_{1,0})$ rate 
are normalized to the corresponding experimental $E(2^+_1)$ 
excitation energy 
and $B(E2;2^+_{1}\to0^+_{1})$ value \cite{rajbanshi2021}, respectively.}
\label{fig:kr82}
\end{center}
\end{figure*}

\begin{figure*}[ht]
\begin{center}
\includegraphics[width=\linewidth]{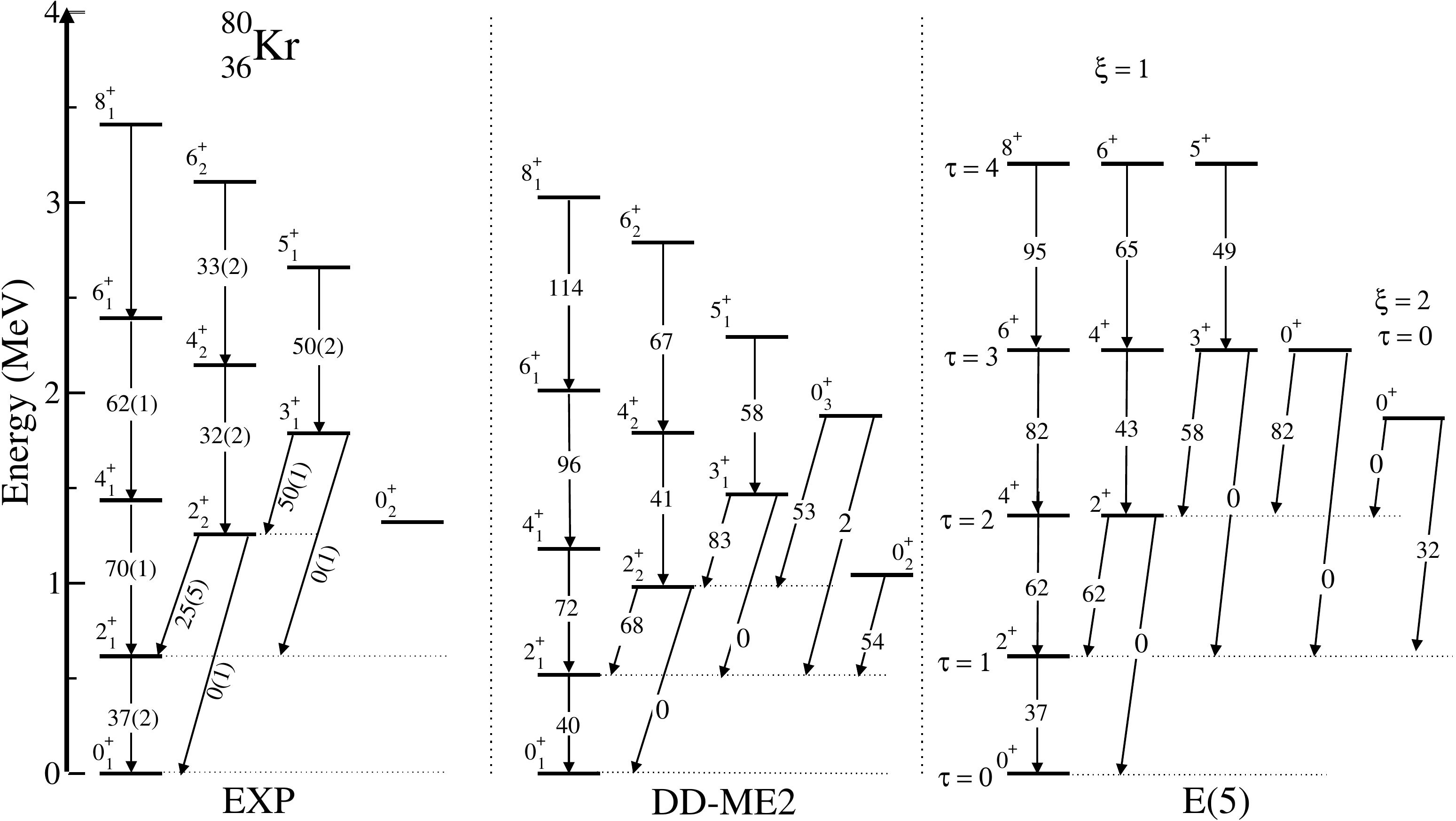}
\caption{Same as Fig.~\ref{fig:kr82}, but for the $^{80}$Kr nucleus. 
The experimental data are taken from Ref.~\cite{data}.}
\label{fig:kr80}
\end{center}
\end{figure*}

\subsubsection{Detailed level structure\label{sec:spectra}}

Let us look into a more detailed energy-level structure 
of individual nuclei. 
Here we specifically consider the transitional nuclei 
$^{82}$Kr and $^{80}$Kr, since particularly the former has 
been suggested \cite{rajbanshi2021} to be 
a candidate for the E(5) critical-point nucleus. 
The triaxial RHB energy surface for the nucleus $^{82}$Kr 
indeed exhibits a flat-bottomed potential that weakly depends 
on the $\gamma$ deformation (see Fig.~\ref{fig:pes}), 
and that most closely resembles the E(5) potential. 

A well-known fact is that the moments of inertia 
computed by using the IB formula 
are considerably smaller than the empirical values. 
In order to effectively take into account this deficiency, 
the IB moments of inertia have often been increased 
by $\approx30-40$ \% in many of the previous calculations 
using collective Hamiltonian (see, e.g., Ref.~\cite{li2010}). 
In the same spirit, 
and in order to make a meaningful comparison between 
the theoretical and experimental energy spectra, 
we show, in this particular section, 
the spectroscopic results for $^{82}$Kr and $^{80}$Kr 
obtained from the QCH 
calculation with the IB moment of inertia increased by 40 \%. 
The effect of the increase is such that excitation 
energies for all the states, except for the $0^+$ ones, 
are lowered by approximately $10-30$ \%. 

Note also that the $2^+_{\xi=1,\tau=1}$ energy level and 
$B(E2;2^+_{\xi=1,\tau=1}\to0^+_{\xi=1,\tau=0})$ value 
in the E(5) spectrum are normalized to the 
experimental \cite{rajbanshi2021,data} $2^+_1$ levels and 
$B(E2;2^+_{1}\to0^+_{1})$ values, respectively, where 
$\xi$ and $\tau$ are quantum numbers of E(5) 
\cite{iachello2000}.

Figure~\ref{fig:kr82} 
shows the computed excitation spectra and $B(E2)$ transition 
strengths for $^{82}$Kr, obtained from the 
RHB+QCH method that uses the DD-ME2 EDF, in comparison to 
the corresponding experimental and E(5) energy spectra. 
One notices that the present calculation reproduces 
the overall feature of the experimental energy spectrum. 
The calculated $B(E2)$ transition rates within the ground-state, 
$K=0^+_1$ band are generally larger than the experimental values. 
Especially, the calculation gives increasing 
inband $E2$ transition strength as a function of the 
angular momentum within the $K=0^+_1$ band, and this systematic 
trend disagrees with the data but agrees with E(5).

The quasi-$\gamma$, or $K=2^+_\gamma$ band here comprises 
the $2^+_2$, $3^+_1$, $4^+_2$, $5^+_1$, and $6^+_2$ states. 
The energy level of the $2^+_2$ bandhead state 
is predicted to be below that of the $4^+_1$ state, which 
is a typical feature of the $\gamma$-soft nucleus and 
is also consistent with the empirical trend. 
The RHB+QCH result, however, indicates a staggering 
pattern that is characterized by 
the near degeneracy of the even- and odd-spin members 
of the band, i.e., $(3^+_\gamma,4^+_\gamma)$, 
$(5^+_\gamma,6^+_\gamma)$, \ldots. 
This energy-level pattern is consistent with 
the E(5) symmetry, and is characteristic of the 
$\gamma$-unstable rotor model \cite{gsoft}, or, equivalently, 
the O(6) symmetry of the IBM. 
In addition, the calculated $K=2^+_\gamma$ 
band exhibits the $E2$ selection rule for the 
interband transitions to the $K=0^+_1$ band 
in agreement with the experimental data and E(5).

The $0^+_3$ state in the present RHB+QCH calculation is here 
associated with the $0^+_{1,3}$ state of E(5). 
The calculated $0^+_3$ excitation energy is close to 
the experimental data, but is rather lower than the 
corresponding $0^+_{1,3}$ level of E(5). 
The nearly vanishing $B(E2; 0^+_3\to2^+_2)/B(E2;0^+_3\to2^+_1)$ 
branching ratio is here obtained, consistently with the 
experimental data and with the selection rule of E(5).

Our model further predicts the $K=0^+_2$ band consisting of 
the $0^+_2$, $2^+_3$, $4^+_3$, \ldots states, which is associated 
with part of the $\xi=2$ family in the E(5) spectrum. 
An overall feature of the calculated $K=0^+_2$ looks similar 
to that of the experimental counterpart. 
The $B(E2;0^+_2\to2^+_1)$ value is here computed to be 18 W.u., 
in a fair agreement with both the experimental value 
and $B(E2;0^+_{2,0}\to2^+_{1,1})$ rate of E(5). 
However, the calculated $K=0^+_2$ band is considerably lower in energy 
than the observed one and $\xi=2$ band of E(5). 
Especially, the energy level of the bandhead state $0^+_2$ 
is here predicted to be below the $2^+_2$ level, which disagrees 
with the data and E(5). 
A previous five-dimensional collective Hamiltonian 
calculation based on the HFB method using the Gogny D1S EDF 
\cite{delaroche2010} obtained the $0^+_2$ excitation energy for the $^{82}$Kr 
isotopes, that is slightly lower than the $2^+_2$ one. 
In the symmetry-projected triaxial GCM calculation also using 
the Gogny-D1S EDF \cite{trodriguez2014}, the $0^+_2$ 
energy level was predicted to be lower than 
the $2^+_2$ one for $^{82}$Kr. 

\begin{table*}
\caption{
Calculated and experimental energy and $B(E2)$ ratios 
for low-lying states of the 
transitional nuclei $^{82}$Kr and $^{80}$Kr, 
and the corresponding E(5) limits. 
The states are labelled by the E(5) quantum numbers $\xi$ and $\tau$. 
The DD-ME2 EDF is used for the calculation. 
The experimental data for $^{82}$Kr and $^{80}$Kr 
are taken from Refs.~\cite{rajbanshi2021} and \cite{data}, respectively, 
while the E(5) values are taken from \cite{iachello2000}. 
\label{tab:e5}
}
 \begin{center}
 \begin{ruledtabular}
  \begin{tabular}{lccccc}
& \multicolumn{2}{c}{$^{82}$Kr}
& \multicolumn{2}{c}{$^{80}$Kr} & \\
\cline{2-3}
\cline{4-5}
Ratio & Experiment & DD-ME2 &Experiment & DD-ME2 & E(5) \\
\hline
$E(4^+_{1,2})/E(2^+_{1,1})$ &
2.34 & 2.33 & 2.33 & 2.28 & 2.20 \\
$E(0^+_{2,0})/E(2^+_{1,1})$ &
2.80 & 2.16 & 2.14 & 2.01 & 3.03 \\
$E(0^+_{1,3})/E(2^+_{1,1})$ &
3.2 & 3.07 & & 3.63 & 3.59 \\
$E(0^+_{2,0})/E(0^+_{1,3})$ &
0.9 & 0.70 & & 0.55 & 0.84 \\
$E(2^+_{1,2})/E(2^+_{1,1})$ &
1.90 & 2.00 & 2.04 & 1.82 & 2.20 \\
$\frac{B(E2;4^+_{1,2}\to2^+_{1,1})}{B(E2;2^+_{1,1}\to0^+_{1,0})}$ &
$1.48\pm0.16$ & 1.81 & $1.88\pm0.29$ & 1.82 & 1.68 \\
$\frac{B(E2;2^+_{1,2}\to2^+_{1,1})}{B(E2;2^+_{1,1}\to0^+_{1,0})}$ &
$1.67\pm0.30$ & 1.09 & $0.67\pm0.14$ & 1.75 & 1.68 \\
$\frac{B(E2;0^+_{2,0}\to2^+_{1,1})}{B(E2;2^+_{1,1}\to0^+_{1,0})}$ &
$0.57\pm0.10$ & 0.80 & & 1.38 & 0.86 \\
$\frac{B(E2;0^+_{2,0}\to2^+_{1,2})}{B(E2;0^+_{2,0}\to2^+_{1,1})}$ &
$0.2\pm0.3$ & 2.98 & & 0.0020 & 0 \\
$\frac{B(E2;0^+_{1,3}\to2^+_{1,1})}{B(E2;0^+_{1,3}\to2^+_{1,2})}$ &
$0.1\pm0.1$ & 0.22 & & 0.036 & 0 \\
  \end{tabular}
 \end{ruledtabular}
 \end{center}
\end{table*}

In Fig.~\ref{fig:kr80} we make a similar comparison 
for the adjacent nucleus $^{80}$Kr. 
As we can see in Fig.~\ref{fig:pes}, the corresponding 
SCMF energy surface for $^{80}$Kr is almost flat in the 
$\gamma$ direction, which is a manifestation of the O(6) symmetry. 
Figure~\ref{fig:kr80} shows that 
the energy spectrum calculated for $^{80}$Kr by the RHB+QCH model 
is, qualitatively, in a better agreement with the experimental 
data than in the case of $^{82}$Kr. 
The calculated spectrum is, however, generally more 
compressed than the experimental one. 
As compared to the E(5) spectrum, 
both the calculated and experimental $0^+_2$ energy levels 
are much lower than the corresponding $0^+_{2,0}$ one. 
In contrast to $^{82}$Kr, we obtain for $^{80}$Kr 
a weak $E2$ transition, $B(E2;0^+_2\to2^+_2)=0.11$ W.u. 
This conforms to the E(5) selection rule, which forbids 
the $E2$ transition $0^+_{2,0}\to2^+_{1,2}$.

In the third column of Table~\ref{tab:e5} the calculated 
energy and $B(E2)$ ratios of low-lying states 
of the $^{82}$Kr nucleus, obtained with the DD-ME2 EDF, 
are shown.  
The results are compared with the corresponding experimental 
values \cite{rajbanshi2021} (the second column) 
and E(5) limits \cite{iachello2000} (last column). 
In the table, the states are labelled by the E(5) quantum 
numbers as $0^+_{1,0}$, $2^+_{1,1}$, 
$4^+_{1,2}$, $2^+_{1,2}$, $0^+_{2,0}$, and $0^+_{1,3}$, 
which are, respectively, associated with the 
$0^+_1$, $2^+_1$, $4^+_1$, 
$2^+_2$, $0^+_2$, and $0^+_3$ states in the calculation 
as well as the experiment. 
Of particular interest are the properties of the excited $0^+$ states. 
We notice that the ratio $E(0^+_{2,0})/E(2^+_{1,1})$ 
in our calculation is considerably smaller than the 
E(5) value, 3.03. 
The $E2$ selection rule for the $0^+_{1,3}$ state of E(5), 
i.e., 
$\frac{B(E2;0^+_{1,3}\to2^+_{1,1})}{B(E2;0^+_{1,3}\to2^+_{1,2})}=0$ 
is reasonably accounted for in our model. 
On the other hand, we obtain a large branching ratio 
$\frac{B(E2;0^+_{2,0}\to2^+_{1,2})}{B(E2;0^+_{2,0}\to2^+_{1,1})}=2.98$, 
which disagree with the data and with E(5). 
Note that the $0^+_{2,0}\to2^+_{1,2}$ transition 
is here predicted to be particularly strong, 
i.e., $B(E2;0^+_2\to2^+_2)=55$ W.u..

The results for $^{80}$Kr are shown in the fifth column 
of Table~\ref{tab:e5}. 
Both the theoretical and experimental 
$E(0^+_{2,0})/E(2^+_{1,1})$ ratios are considerably 
smaller than the E(5) value. 
The predicted ratio $E(0^+_{1,3})/E(2^+_{1,1})=3.63$ 
for $^{80}$Kr is here suggested to be closer to the E(5) 
value, 3.59, than for $^{82}$Kr. 
It is worthwhile to remark that, 
in accordance with the E(5) selection rules for 
the $E2$ transitions, the present calculation 
gives nearly vanishing values for both the 
$\frac{B(E2;0^+_{2,0}\to2^+_{1,2})}{B(E2;0^+_{2,0}\to2^+_{1,1})}$
and 
$\frac{B(E2;0^+_{1,3}\to2^+_{1,1})}{B(E2;0^+_{1,3}\to2^+_{1,2})}$
ratios. The result for the former branching ratio is in 
a marked contrast to the one for $^{82}$Kr. 
Nevertheless, since the experimental information about the low-lying 
states of $^{80}$Kr is not as abundant as 
for $^{82}$Kr, an extensive comparison between the RHB+QCH 
result and experiment is difficult.

\section{Concluding remarks\label{sec:summary}}

Based on the framework of the nuclear density functional theory, 
we have investigated the spectroscopic properties that signal the 
shape-phase transitions 
in the chain of the Kr isotopes in the mass $A\approx80$ region, 
with a particular focus on the $^{82}$Kr nucleus, 
which was recently identified as empirical evidence for the E(5) CPS. 
The constrained SCMF calculations within 
the RHB method using two representative classes of the relativistic 
EDF and a pairing interaction have been performed for the 
even-even nuclei $^{76-86}$Kr. 
The SCMF solutions have been then used as the microscopic 
inputs to determine the ingredients of the 
five-dimensional quadrupole collective Hamiltonian, that is, 
the deformation-dependent moments of inertia and mass parameters, 
and the collective potential. 
The diagonalization of the QCH has yielded excitation spectra 
and transition probabilities of the considered Kr nuclei. 

The resultant triaxial quadrupole deformation energy surfaces 
have indicated an evolution of the equilibrium shape as a function 
of the nucleon number (Fig.~\ref{fig:pes}): 
a competition among a nearly spherical global, 
an oblate, and a strongly prolate deformed local 
minima in $^{76,78}$Kr, 
a notable $\gamma$-softness in $^{80}$Kr, 
a weakly-deformed prolate shape characterized by a 
flat-bottomed potential that is soft both in the 
$\beta$ and $\gamma$ deformations for $^{82,84}$Kr, 
and a nearly spherical shape for $^{86}$Kr, corresponding 
to the neutron major shell closure $N=50$. 
The RHB+QCH calculation has provided a reasonable description 
of the experimental low-energy spectra (Fig.~\ref{fig:level}) 
for $N \leq 44$, but overestimates the data for $N \geq 46$ 
as the neutron magic number $N=50$ is approached. 
Around the shell closure, 
the QCH approach, which produces purely 
collective states, is not expected to give a very good 
description of the empirical data. 
The calculated $B(E2)$ rates (Fig.~\ref{fig:e2}) have been shown 
to be generally in agreement with the data, 
whereas particularly the $2^+_2\to2^+_1$ transition 
strengths for $N \leq 44$ have been overestimated, 
due to the strong shape mixing.

The behaviors of the calculated energy and $B(E2)$ ratios, 
and fluctuations in $\beta$ and $\gamma$ deformations along 
the isotopic chain have indicated the underlying 
nuclear structural change around $^{82}$Kr, characterized by 
the significant amount of shape mixing. 
The detailed analyses of the calculated low-energy spectra of the 
transitional nuclei $^{82}$Kr and $^{80}$Kr have been made 
in comparison to the experimental and E(5) spectra 
(cf. Figs.~\ref{fig:kr82} and \ref{fig:kr80}). 
Particularly for $^{82}$Kr, the predicted 
quasi-$\gamma$, $K=2^+_\gamma$ band has been shown to be higher 
than the experimental one, in such a way that the bandhead $2^+_2$ level 
is close in energy to the $4^+_1$ one in the ground-state band. 
Another notable deviation from  the 
experiment as well as from the E(5) symmetry appears in the description 
of the $0^+_2$ energy level, which is here calculated to be so low as to 
be below the $2^+_2$ one for $^{82}$Kr, 
and which shows the $E2$ branching ratio that is quite at variance 
with the experimental data and E(5). 
The deviation from the data has arisen in part from 
the particular choice of the EDFs, which may further 
point to some deficiencies 
of the model when it is applied to this particular mass region.
Another possibility consists in the fact that the employed 
RHB+QCH approach in its current version 
presents a relatively simple model, which is built on the 
triaxial quadrupole shape degrees of freedom only, and thus  
the inclusions of some additional collective degrees of freedom 
in the Hamiltonian, in a similar spirit, e.g., 
to Refs.~\cite{xiang2020,nomura2020pv,nomura2021pv}, 
may improve the description of the data. 
It is an interesting future study to investigate 
these possibilities. 

In conclusion, the RHB+QCH method has demonstrated an ability 
to provide the spectroscopic observables that can be 
directly comparable to the experimental data, based solely on 
a choice of the universal EDF and pairing interaction. 
The approach allows for a timely, systematic and 
computationally feasible theoretical prediction for the nuclear 
shape-related phenomena that are experimentally of much interest, 
such as the shape QPTs and coexistence, and is expected to serve as 
a useful benchmark for more complicated microscopic calculations.

\acknowledgments
The authors are grateful to Antonio Bjel\v ci\'c for helping 
them with implementation of the modified RHB solver. 
This work is financed within the Tenure Track Pilot Programme of 
the Croatian Science Foundation and the \'Ecole Polytechnique 
F\'ed\'erale de Lausanne, and the Project TTP-2018-07-3554 Exotic 
Nuclear Structure and Dynamics, with funds of the Croatian-Swiss 
Research Programme.

\bibliography{refs}

\end{document}